\documentclass[10pt,a4paper,openright,tablecaptionabove]{scrartcl}
\usepackage{subfig}
\usepackage[english]{babel}
\usepackage{amsmath, amsthm, amssymb}
\usepackage{bbm}
\usepackage{url}
\usepackage[colorlinks=true,citecolor = purple,urlcolor=purple, linkcolor=blue,
filecolor=magenta]{hyperref}
\usepackage{tikz-cd}
\usepackage{braket}
\usepackage{path}
\usepackage{nicefrac}
\usepackage{amssymb}
\usepackage{amsmath}
\usepackage{amsthm}
\usepackage{cancel}
\usepackage{indentfirst}
\usepackage{eucal}
\usepackage{appendix}
\usepackage[utf8]{inputenc}
\usepackage{geometry}
\usepackage{verbatim}
\usepackage{nameref}
\usepackage{alltt}
\usepackage{hyperref}
\usepackage{mathdots}
\usepackage{pgfplots}
\usepackage{dsfont}
\usepackage{appendix}
\usepackage[font=small,labelfont=bf]{caption} 

\newtheorem{rem}{Remark}

\newcommand{\figref}[1]{\figurename~\ref{#1}}

\usetikzlibrary{arrows.meta,automata,positioning,shadows}

\usetikzlibrary{decorations.markings}

\areaset[current]{500pt}{680pt}
\numberwithin{equation}{section}

\begin{document}

\thispagestyle{empty}
\begin{center}
\Large{{\bf Polynomial algebras of superintegrable systems separating in Cartesian coordinates from higher order ladder operators}}
\end{center}
\vskip 0.5cm
\begin{center}
\textsc{Danilo Latini, Ian Marquette and Yao-Zhong Zhang}
\end{center}
\begin{center}
	
School of Mathematics and Physics, The University of Queensland, Brisbane, QLD 4072, Australia
\end{center}
\begin{center}
 \footnotesize{d.latini@uq.edu.au, i.marquette@uq.edu.au, yzz@maths.uq.edu.au}
\end{center}
\vskip 0.5cm

\vskip  1cm
\hrule
\begin{center}
\textsf{{\bf abstract}}
\end{center}
 \begin{abstract}
 	\noindent We introduce the general polynomial algebras characterizing a class of higher order superintegrable systems that separate in Cartesian coordinates. The construction relies on underlying polynomial Heisenberg algebras and their defining higher order ladder operators. One feature of these algebras is that they preserve by construction some aspects of the structure of the $\mathfrak{gl}(n)$ Lie algebra. Among the classes of Hamiltonians arising in this framework are various deformations of harmonic oscillator and singular oscillator related to exceptional orthogonal polynomials and even Painlev\'e and higher order Painlev\'e analogs. As an explicit example, we investigate a new three-dimensional superintegrable system related to Hermite exceptional orthogonal polynomials of type III. Among the main results is the determination of the degeneracies of the model in terms of the finite-dimensional irreducible representations of the polynomial algebra. 
 	\end{abstract}
\vskip 0.35cm
\hrule

%
%
%
%
%
\section{Introduction}
\label{sec1}

\noindent Over the past years, it has been demonstrated how the study of superintegrable systems is intimately connected with the study of various algebraic structures beyond Lie algebras \cite{Granovskii1992,Bonatsos:1994fi,PhysRevA.50.3700, BONATSOS1999537}, which take the form of quadratic \cite{doi:10.1063/1.1348026, Post2011, MillerPostWinternitz2013R, Kalnins2013} and more generally polynomial algebras \cite{Marquette2009,Marquette_2010,Kalnins2011,Marquette_2011,EscobarRuiz2017,Marquette2020}. This is due to the fact that integrals are partial differential operators of second or  higher order. 

Recently it has been shown that quadratic algebras associated with $n$-dimensional systems are in general of higher rank \cite{Hoque_2015, Chen_2019, bie2020racah,latini2020embedding}.  These algebraic structures allow one to obtain useful information on quantum systems and their degenerate spectrum \cite{2021_correa}.  In most cases, they do not display an obvious basis and thus the construction of their representations is difficult.

Similarly to the two-dimensional anisotropic oscillator \cite{Bonatsos:1993st} and related symmetry algebra, many papers were devoted to two-dimensional superintegrable systems \cite{Post2012, doi:10.1063/1.4798807,doi:10.1063/1.4823771,Marquette2014}  with deformations for which the wavefunctions were written in terms of exceptional orthogonal polynomials \cite{GmezUllate2009, GmezUllate2010, Odake2009,Odake2010, Quesne2008, Quesne2009, doi:10.1063/1.3671966, Grandati2011, Grandati2013, GomezUllate2004, Carinena_2008, Fellows_2009, doi:10.1063/1.3651222, GRANDATI20122411}. 
Since of their first appearance, exceptional orthogonal polynomials have found application in many different physical contexts \cite{doi:10.1063/1.166056, doi:10.1142/S0217732396001557, HOQUE2018203, HOFF, 2019plu, 2019plu2,  ChalGru}. One class of Hamiltonians with richer structures such as different \textquotedblleft regimes\textquotedblright for  degeneracies were discovered, leading to the path of different types of spectral design for the quantum systems. It was found how  polynomial algebras of arbitrary degree with three generators allow complicated patterns for the irreducible representations (irreps) connected with the degeneracies \cite{doi:10.1063/1.4798807, doi:10.1063/1.4823771}. These works also allowed one to describe an important feature of the irreps of the polynomial algebras for which, so far, only very limited results have been obtained.

This paper is devoted to introduce the general polynomial algebras for models allowing separation of variables in Cartesian coordinates whose ladder operators can be of arbitrary order. These polynomial algebras, whose structures share some similarities with the well-known $\mathfrak{gl}(n)$ Lie algebra, are very interesting from the point of view of representations, as they allow one to rely on analog of using Cartan-like and ladder-like generators. 

We want to address that these quantum models with underlying polynomial Heisenberg algebras \cite{C_1999, Carballo_2004}, even though they possess separation of variables in Cartesian coordinates, have in general complicated expressions for their wavefunctions beyond hypergeometric type of orthogonal polynomials. More importantly, the patterns for their degenerate spectrum are non-trivial and the separated equations do not provide information on the structures of multiplets. In this paper, we obtain interesting patterns for finite-dimensional unitary representations of the polynomial algebra and use them to provide further insights into the degeneracies of the models. In particular, for the system associated with type III Hermite exceptional orthogonal polynomials, we find that the representations of the polynomial algebra display similarities with the $\mathfrak{su}(3)$ representations, which we will refer to as $\mathfrak{su}(3)$-like.  We obtain direct sums of such representations and the dimensions and multiplicities of the representations.

The paper is organized as follows. In Section \ref{sec2} we construct the higher order polynomial algebra characterizing $n$D superintegrable Hamiltonian systems separating in Cartesian coordinates and whose constituents ladder operators, together with their associated one-dimensional Hamiltonians,  define polynomial Heisenberg algebras.  In Section \ref{sec3} we restrict to the three-dimensional case. Specifically, we study a new three-dimensional superintegrable Hamiltonian system that saparates in Cartesian coordinates $(x_1, x_2, x_3)$. This model, that we introduce in Section \ref{ssec3.1},  due to the presence of a rationally extended oscillator in the $x_1$ axis, turns out to be characterized by a separable solution arising as the product of two standard Hermite orthogonal polynomials (in the $x_2, x_3$ axes) and Hermite exceptional orthogonal polynomials of type III (in the $x_1$ axis). As a direct consequence, the model presents complicated patterns for the degeneracies of the energy levels, that we aim to characterize in terms of finite-dimensional irreps of the polynomial algebra introduced in the previous Section. Finally,  Section \ref{conrem} is devoted to  some concluding remarks.

\section{From polynomial Heisenberg algebras to polynomial symmetry algebras}

\label{sec2}

\noindent Let us consider a $n$-dimensional  Hamiltonian system admitting separation of variables in Cartesian coordinates:

\begin{equation}
\hat{H}=\sum_{i=1}^n \hat{H}_i=\sum_{i=1}^n \bigl(-\partial_{x_ix_i}+V(x_i)\bigl)
\label{eqsep}
\end{equation}

\noindent and let us assume the existence of $n$ pairs of ladder operators $(\hat{A}_i, \hat{A}_i^\dagger)$ of order $k_i$, $i=1, \dots, n$, satisfying the following Polynomial Heisenberg Algebras (PHA) \cite{C_1999, Carballo_2004}:
\begin{align}
[\hat{H}_i, \hat{A}_i]&=-\alpha_i \hat{A}_i \qquad  [\hat{H}_i, \hat{A}^{\dagger}_i]=\alpha_i \hat{A}^\dagger_i \label{eq:ladderop1} \\
\hat{A}_i^{\dagger}\hat{A}_i&=\phi_i(\hat{H}_i) \qquad \quad \hat{A}_i\hat{A}_i^\dagger=\phi_i(\hat{H}_i+\alpha_i)  \, ,
\label{eq:ladderop2}
\end{align}
where $\alpha_i$ are constants and $\phi_i(\hat{H}_i)$ are $n$ polynomials of order $k_i$ in $\hat{H}_i$ . \noindent Clearly, by virtue of \eqref{eq:ladderop2}, we can rewrite the commutators among the ladder operators as:
\begin{equation}
[\hat{A}_i, \hat{A}_i^\dagger]=\phi_i(\hat{H}_i+\alpha_i)-\phi_i(\hat{H}_i) \equiv P_i(\hat{H}_i)\, ,
\label{eq:PHA}
\end{equation}
where $P_i(\hat{H}_i)$ are $n$ polynomials of order $k_i-1$. We also notice that \eqref{eq:ladderop2} implies:
\begin{equation}
\phi_i(\hat{H}_i+\alpha_i)\hat{A}_i=\hat{A}_i\phi_i(\hat{H}_i) \, , \qquad \phi_i(\hat{H}_i)\hat{A}^\dagger_i=\hat{A}^\dagger_i\phi_i(\hat{H}_i+\alpha_i) \, .
\label{eq:imp}
\end{equation}
\noindent Moreover, the following additional relations hold:
\begin{align}
 [\hat{H_i}, (\hat{A}_i^{\dagger})^{s_i}]&=\alpha_i s_i (\hat{A}_i^{\dagger})^{s_i} \, , \qquad	[\hat{H_i}, (\hat{A}_i)^{s_i}]=-\alpha_i {s_i}(\hat{A}_i)^{s_i} \label{eq:operators1}\\
(\hat{A}_i^\dagger)^{s_i}(\hat{A}_i)^{s_i}&=\Phi_i(s_i,\hat{H}_i) \, , \quad (\hat{A}_i)^{s_i}(\hat{A}_i^\dagger)^{s_i}=\Phi_i({s_i},\hat{H}_i+\alpha_i s_i) \, ,
\label{eq:operators}
\end{align}
for $s_i \in \mathds{N}^*$, where we defined:
\begin{equation}
\Phi_i(s_i, x_i):=\prod_{j=1}^{s_i} \phi_i\bigl(x_i-\alpha_i(s_i-j)\bigl) \, .
\label{eq:fi}
\end{equation}
\noindent Proofs can be obtained by induction. For example, if we take the first equation in \eqref{eq:operators}, then for $s_i=1$ is trivially satisfied as the product collapses to $\phi_i(x_i)$. Now, suppose that for $s_i-1  \geq 1$ is satisfied, then:

\begin{align}
(\hat{A}_i^\dagger)^{s_i}(\hat{A}_i)^{s_i}=\hat{A}_i^\dagger (\hat{A}_i^\dagger)^{s_i-1}(\hat{A}_i)^{s_i-1}\hat{A}_i=A^\dagger_i \biggl(\prod_{j=1}^{s_i-1} \phi_i\bigl(x_i-\alpha_i(s_i-j-1)\bigl)\biggl)A_i &\overset{\eqref{eq:imp}}{=}\biggl(\prod_{j=1}^{s_i-1} \phi_i\bigl(x_i-\alpha_i(s_i-j)\bigl)\biggl)A^\dagger_i A_i \nonumber \\
&\overset{\eqref{eq:ladderop2}}=\biggl(\prod_{j=1}^{s_i-1}\phi_i\bigl(x_i-\alpha_i(s_i-j)\bigl)\biggl)\phi_i(x_i) \nonumber\\
&\,\,=\prod_{j=1}^{s_i} \phi_i\bigl(x_i-\alpha_i(s_i-j)\bigl) \nonumber \, ,
\end{align} 
which is the desired result. The other relation can be obtained in a similar way. We notice that an equivalent form for the product  \eqref{eq:fi} also exists, as the following equality holds:
\begin{equation}
\prod_{j=1}^{s_i} \phi_i\bigl(x_i-\alpha_i(s_i-j)\bigl)=\prod_{j=1}^{s_i} \phi_i\bigl(x_i-\alpha_i(j-1)\bigl) \, .
\label{equality}
\end{equation}
 This can be shown by expanding the two products:
 \begin{equation}
 \phi_i(x_i-\alpha_i(s_i-1))\phi_i(x_i-\alpha_i(s_i-2)) \dots \phi_i(x_i-\alpha_i)\phi_i(x_i) = \phi_i(x_i) \phi_i(x_i-\alpha_i)\dots \phi_i(x_i-\alpha_i(s_i-1))  \, .
 \label{proofs}
 \end{equation}
\noindent At this point, let us introduce the following $n(n-1)$ operators of order $s_i k_i+s_j k_j$:
\begin{align}
\hat{E}_{ij}&:=(\hat{A}^{\dagger}_i)^{s_i} (\hat{A}_j)^{s_j} \qquad (i \neq j) \, .
\label{eq:constants}
\end{align}
If we require that they commute with the Hamiltonian \eqref{eqsep}, we get:
\begin{equation}
[\hat{H}, \hat{E}_{ij}]
=(\alpha_i s_i-\alpha_j s_j)\hat{E}_{ij} \, .
\label{comm2}
\end{equation}

\noindent This leads us to the condition: $\alpha_i s_i=\alpha_j s_j$, with $\alpha_1 s_1=\alpha_2 s_2=\dots =\alpha_n s_n=:\alpha$. Together with the above operators, we consider the one-dimensional Hamiltonians $\hat{H}_i$ ($i=1, \dots, n$). The set composed by the $n^2$ operators $\{\hat{E}_{ij}, \hat{H}_i\}$ close in the following higher rank polynomial algebra:
\begin{align}
[\hat{E}_{ij}, \hat{E}_{k\ell}] &= \delta_{jk}\bigl(\Phi_k({s_k},\hat{H}_k+\alpha)-\Phi_i({s_k},\hat{H}_k)\bigl)\hat{E}_{i\ell}- \delta_{i\ell}\bigl(\Phi_\ell({s_\ell},\hat{H}_i+\alpha)-\Phi_i({s_\ell},\hat{H}_\ell)\bigl)\hat{E}_{kj} \label{alg1}\\
[\hat{E}_{ij}, \hat{E}_{j i}] &= \Phi_i(s_i, \hat{H}_i)\Phi_j(s_j, \hat{H}_j+\alpha)-\Phi_i(s_i, \hat{H}_i+\alpha)\Phi_j(s_j, \hat{H}_j) \label{alg2}\\
[\hat{E}_{ij}, \hat{H}_{k}] &=\alpha(\delta_{jk} \hat{E}_{ik}-\delta_{ik} \hat{E}_{kj} ) \, ,
\label{eq:alg}
\end{align}
where the functions $\Phi(s,x)$ are the ones given in \eqref{eq:fi}.
\begin{rem}
When we restrict to the $n$-dimensional harmonic oscillator:
\begin{equation}
\hat{H}=\frac{1}{2}\sum_{i=1}^n \bigl(-\partial_{x_ix_i}+x^2_i\bigl)\, ,
\label{ho}
\end{equation}
the ladder operators are of the first-order $($$k_1=k_2= \dots =k_n=1$$)$. Explicitly:
\begin{equation}
\quad \hat{A}_i=\frac{\partial_{x_i}+x_i }{\sqrt{2}} \, , \quad \hat{A}^\dagger_i=\frac{-\partial_{x_i}+x_i}{\sqrt{2}} 
\label{eq:ladderho}
\end{equation}
and the following commutation relations are satisfied:
\begin{align}
[\hat{H}_i, \hat{A}_i]&=- \hat{A}_i \hskip 1.5cm [\hat{H}_i, \hat{A}^{\dagger}_i]= \hat{A}^\dagger_i \label{eq:ladderoposc} \\
\hat{A}_i^{\dagger}\hat{A}_i&=\hat{H}_i-1/2 \qquad \quad \,\,\hat{A}_i\hat{A}_i^\dagger=\hat{H}_i+1/2  \, .
\label{eq:ladderoposc2}
\end{align}
Thus, in this limiting case we have $s_1=s_2=\dots=s_n$, and for $s_1=s_2=\dots=s_n=1$ we get:
\begin{align}
[\hat{E}_{ij}, \hat{E}_{k\ell}] &= \delta_{jk}{E}_{i\ell}- \delta_{i\ell}\hat{E}_{kj}\\
[\hat{E}_{ij}, \hat{E}_{j i}] &= \hat{H}_i-\hat{H}_j\\
[\hat{E}_{ij}, \hat{H}_{k}] &=\delta_{jk} \hat{E}_{ik}-\delta_{ik} \hat{E}_{kj}  \, ,
\label{eq:algosc}
\end{align}
as $\Phi_i(1, \hat{H}_i) \equiv \phi_i(\hat{H}_i)=\hat{H}_i-1/2$ and $\alpha=1 \, , \forall \, i=1, \dots, n$.
Notice also that if we rewrite $\hat{H}_k\equiv \hat{E}_{kk}$ $($i.e. we allow equal indices for the generators \eqref{eq:constants}$)$ we can recast the algebra in the compact form:
\begin{align}
[\hat{E}_{ij}, \hat{E}_{k\ell}] &= \delta_{jk}\hat{E}_{i\ell}- \delta_{i\ell}\hat{E}_{kj} \quad (i,j=1, \dots, n) \, ,
\end{align}
which gives the usual realisation of the $\mathfrak{u}(n)$ Lie algebra for the $n$-dimensional harmonic oscillator. 
\end{rem}
\noindent The $n=2$ case has been already studied in the literature (see \cite{doi:10.1063/1.4798807, doi:10.1063/1.4823771} and references therein). The Hamiltonian is:
\begin{equation}
\hat{H}=\hat{H}_1+\hat{H}_2
\label{ham}
\end{equation}
\noindent and the four operators $\{\hat{H}_1, \hat{H}_2,  \hat{E}_{12}, \hat{E}_{21}\}$ are all conserved when $\alpha_1 s_1=\alpha_2 s_2=:\alpha$, with commutation relations:
\begin{align}
[\hat{E}_{12}, \hat{E}_{21}] &= \Phi_1(s_1, \hat{H}_1)\Phi_2(s_2, \hat{H}_2+\alpha)-\Phi_1(s_1, \hat{H}_1+\alpha)\Phi_2(s_2, \hat{H}_2)\\
[\hat{E}_{12}, \hat{H}_{1}] &=-\alpha \hat{E}_{12}  \qquad 
[\hat{E}_{12}, \hat{H}_{2}] =\alpha \hat{E}_{12}  \\
[\hat{E}_{21}, \hat{H}_{1}] &=\alpha \hat{E}_{21}  \hskip 1cm
[\hat{E}_{21}, \hat{H}_{2}] =-\alpha \hat{E}_{21}  \, .
\label{eq:u2}
\end{align}

\noindent By considering the following combinations of the generators:
\begin{equation}
\hat{T}_3 :=\frac{\hat{H}_1-\hat{H}_2}{2 \alpha} \qquad \hat{T}_{+}:=\hat{E}_{12}  \qquad \hat{T}_{-}:=\hat{E}_{21}  \, ,
\label{eq:defu2}
\end{equation}
together with $[\hat{H},\cdot]=0$ we get:
\begin{align}
[\hat{T}_{+}, \hat{T}_{-}] &= \Phi_1(s_1, \hat{H}/2+\alpha \hat{T}_3)\Phi_2(s_2, \hat{H}/2-\alpha \hat{T}_3+\alpha)-\Phi_1(s_1, \hat{H}/2+\alpha \hat{T}_3+\alpha)\Phi_2(s_2, \hat{H}/2-\alpha \hat{T}_3)\\
[\hat{T}_3, \hat{T}_{\pm}] &=\pm T_{\pm} \, .
\label{eq:eu2}
\end{align}
\noindent In this way, we restrict to the set of operators $\{\hat{H}, \hat{T}_3,\hat{T}_\pm\}$. At this point, by introducing the new function: 
\begin{equation}
\Xi^{(s_1, s_2)}_{12}(\hat{H},\hat{T}_3):=\Phi_1(s_1, \hat{H}/2+\alpha \hat{T}_3)\Phi_2(s_2, \hat{H}/2-\alpha \hat{T}_3+\alpha)
\end{equation}
we can recast the polynomial algebra in the following form:
\begin{align}
 [\hat{H},\cdot]&=0 \label{eq:eu2s0} \\
[\hat{T}_{+}, \hat{T}_{-}] &= \Xi_{12}^{(s_1, s_2)}(\hat{H},\hat{T}_3)-\Xi_{12}^{(s_1, s_2)}(\hat{H},\hat{T}_3+1) \label{eq:eu2s1}\\
[\hat{T}_3, \hat{T}_{\pm}] &=\pm \hat{T}_{\pm} \, .
\label{eq:eu2s3}
\end{align}
\noindent This polynomial algebra is of order $s_1  + s_2 -1$ in the generators ($k_1 s_1+k_2 s_2-1$ in terms of differential operators).

\begin{rem}
When we restrict to the $2$D harmonic oscillator:
 $$s_1 =s_2=\alpha=1$$ $$\Xi^{(s_1, s_2)}_{12}(\hat{H},\hat{T}_3)-\Xi_{12}^{(s_1, s_2)}(\hat{H},\hat{T}_3+1)= 2 \hat{T}_3$$ 
 
\noindent the polynomial algebra collapses to:
\begin{align*}
[\hat{H},\cdot]=0 \qquad
[\hat{T}_{+}, \hat{T}_{-}] = 2 \hat{T}_3 \qquad
[\hat{T}_3, \hat{T}_{\pm}] =\pm \hat{T}_{\pm} \, ,
\end{align*}
with $\hat{T}_+=\hat{A}_1^\dagger \hat{A}_2$ and  $\hat{T}_-=\hat{A}_2^\dagger \hat{A}_1$, where $(\hat{A}_i,\hat{A}_i^\dagger)$ are the usual first-order ladder operators defined in \eqref{eq:ladderho}.
\end{rem}
\noindent 
The first non-trivial example is represented by the $2$D anisotropic oscillator, which has been investigated in \cite{Bonatsos:1993st} (see  \cite{Bonatsos:1994fi} for the general $n$D case). The Hamiltonian and ladder operators read \cite{Bonatsos:1993st}:
\begin{equation}
\hat{H}=\hat{H}_1+\hat{H}_2=\frac{1}{2} \bigl(-\partial_{x_1x_1}+x^2_1/m_1^2\bigl)+\frac{1}{2} \bigl(-\partial_{x_2x_2}+x^2_2/m_2^2\bigl)\, ,
 \quad \hat{A}^\dagger_i=\frac{-\partial_{x_i}+x_i/m_i}{\sqrt{2}} \, , \quad \hat{A}_i=\frac{\partial_{x_i}+x_i/m_i }{\sqrt{2}}
 \label{hamaniladders} 
\end{equation}
where $m_1, m_2$ are mutually prime natural numbers. The above operators close the commutation relations ($i=1,2$):
\begin{align}
[\hat{H}_i, \hat{A}_i]&=- \hat{A}_i/m_i \hskip 1.65cm [\hat{H}_i, \hat{A}^{\dagger}_i]= \hat{A}^\dagger_i/m_i \label{eq:ladderani1} \\
\hat{A}_i^{\dagger}\hat{A}_i&=\hat{H}_i-1/(2m_i) \qquad \quad \,\,\,\hat{A}_i\hat{A}_i^\dagger=\hat{H}_i+1/(2m_i)  \, .
\label{eq:ladderani2}
\end{align}
\noindent The two operators $\hat{E}_{12}$, $\hat{E}_{21}$ commute with the Hamiltonian $\hat{H}$ if $s_1/m_1=s_2/m_2=\alpha$, i.e. for $s_i=m_i$ and $\alpha=1$. Thus, the generators of the polynomial algebra (of order $m_1+m_2-1$) turn out to be:
\begin{equation}
\hat{H}=\hat{H}_1+\hat{H}_2 \, , \quad \hat{T}_3:=\frac{\hat{H}_1-\hat{H}_2}{2} \, ,\quad \hat{T}_{+}:=\hat{E}_{12}=(A_1^\dagger)^{m_1}(A_2)^{m_2} \, , \quad \hat{T}_{-}:=\hat{E}_{21}=(A_2^\dagger)^{m_2}(A_1)^{m_1} \, ,
\label{eq:defu2ani}
\end{equation}
and the higher order polynomial relations \eqref{eq:eu2s1}-\eqref{eq:eu2s3} are obtained through the following structure function:
\begin{equation}
\small \Xi_{12}^{(m_1, m_2)}(\hat{H},\hat{T}_3):=\Phi_1(m_1, \hat{H}/2+ \hat{T}_3)\Phi_2(m_2, \hat{H}/2- \hat{T}_3+1)=\prod_{i=1}^{m_1} \biggl(\hat{H}/2+\hat{T}_3-\frac{2i-1}{2m_1}\biggl)\prod_{j=1}^{m_2} \biggl(\hat{H}/2-\hat{T}_3+\frac{2 j-1}{2m_2}\biggl) \, .
\label{eq:funanis}
\end{equation}

\noindent When $m_1=m_2=1$ the $\mathfrak{u}(2)$ algebra is recovered. Thus, the polynomial algebra \eqref{eq:eu2s0}-\eqref{eq:eu2s3} characterized by the function \eqref{eq:funanis} represents a non-linear generalization of the $\mathfrak{u}(2)$ algebra. The finite-dimensional irreps of this algebra have been constructed by using a deformed oscillator algebra approach \cite{PhysRevA.50.3700, BONATSOS1999537, doi:10.1063/1.1348026}. The same strategy has been in used in \cite{doi:10.1063/1.4823771} to obtain finite-dimensional irreps of the polynomial algebras describing 2D superintegrable systems separable in Cartesian coordinates related to a rational extension of the harmonic oscillator with type III Hermite exceptional orthogonal polynomials. Many other 2D superintegrable systems have been studied within this framework (see \cite{Marquette_2010, Marquette_2011, Marquette2019} and references therein).

\section{Three-dimensional superintegrable Hamiltonian systems}
\label{sec3}

\noindent In this Section \ref{sec3}, we are interested in constructing polynomial algebras for $3$D superintegrable systems separating in Cartesian coordinates starting from polynomial Heisenberg algebras \eqref{eq:ladderop1}-\eqref{eq:ladderop2}. The $3$D Hamiltonian reads:
\begin{equation}
\hat{H}=\hat{H}_1 +\hat{H}_2+\hat{H}_3
\label{ham3D}
\end{equation}

\noindent and we can introduce the  following set composed by the nine operators:
\begin{equation}
\{\hat{H}_1, \hat{H}_2, \hat{H}_3, \hat{E}_{12}, \hat{E}_{13}, \hat{E}_{21}, \hat{E}_{23}, \hat{E}_{31}, \hat{E}_{32}\} \, ,
\end{equation}

\noindent satisfying the condition $\alpha_1 s_1=\alpha_2 s_2=\alpha_3 s_3=\alpha$. The nonzero commutation relations are obtained from \eqref{alg1}-\eqref{eq:alg} and,
in complete analogy with the $2$D case, we consider the following combinations:
\begin{align}
&\hat{T}_3:=\frac{\hat{H}_1-\hat{H}_2}{2\alpha} \qquad \hat{Y}:=\frac{\hat{H}_1+\hat{H}_2-2\hat{H}_3}{3 \alpha}\, ,\label{cartan}\\
& \hat{T}_{+}:=\hat{E}_{12} \, , \quad \hat{T}_{-}:=\hat{E}_{21} \, ,\quad  \hat{U}_{+}:=\hat{E}_{23} \, , \quad \hat{U}_{-}:=\hat{E}_{32} \quad \hat{V}_{+}:=\hat{E}_{13} \, , \quad \hat{V}_{-}:=\hat{E}_{31}\, .
\label{eq:su(3)}
\end{align}
In this way, we restrict to the set of operators $\{\hat{H}, \hat{T}_3,\hat{Y}, \hat{T}_\pm, \hat{U}_{\pm}, \hat{V}_\pm\}$.  Besides $[\hat{H}, \cdot]=0$, these operators close under commutation to give the  polynomial algebra:
\begin{equation}
[\hat{T}_3,\hat{Y}]=0
\label{polcar}
\end{equation}
\begin{align}
[\hat{T}_{+},\hat{T}_-]&=\Xi^{(s_1, s_2)}_{12}(\hat{H},\hat{T}_3,\hat{Y})-\Xi_{1 2}^{(s_1, s_2)}(\hat{H},\hat{T}_3+1,\hat{Y}) \hskip 2cm [\hat{T}_3,\hat{T}_\pm]=\pm\hat{T}_\pm  \hskip 1.2 cm [\hat{Y},\hat{T}_\pm]=0 \label{poly1}\\
[\hat{V}_{+},\hat{V}_-]&=\Xi_{1 3}^{(s_1, s_3)}(\hat{H},\hat{T}_3,\hat{Y})-\Xi_{13}^{(s_1, s_3)}(\hat{H},\hat{T}_3+1/2,\hat{Y}+1) \hskip 1.05cm [\hat{T}_3,\hat{V}_\pm]=\pm\frac{1}{2}\hat{V}_\pm  \hskip 0.95cm [\hat{Y},\hat{V}_\pm]=\pm\hat{V}_\pm\\
[\hat{U}_{+},\hat{U}_-]&=\Xi_{2 3}^{(s_2, s_3)}(\hat{H},\hat{T}_3,\hat{Y})-\Xi_{2 3}^{(s_2, s_3)}(\hat{H},\hat{T}_3-1/2,\hat{Y}+1) \hskip 1.05cm [\hat{T}_3,\hat{U}_\pm]=\mp\frac{1}{2}\hat{U}_\pm \hskip 0.85 cm [\hat{Y},\hat{U}_\pm]=\pm \hat{U}_\pm \,   \\
& [\hat{T}_\pm, \hat{V}_\mp]=\mp \bigl(\Phi_1(s_1, \hat{H}/3 +\alpha \hat{T}_3+\alpha \hat{Y}/2+\alpha)-\Phi_1(s_1; \hat{H}/3 +\alpha \hat{T}_3+\alpha \hat{Y}/2)\bigl)\hat{U}_\mp \\    
&[\hat{T}_{\pm},\hat{U}_\pm]=\pm \bigl(\Phi_2(s_2, \hat{H}/3 -\alpha \hat{T}_3+\alpha \hat{Y}/2+\alpha)-\Phi_2(s_2, \hat{H}/3 -\alpha \hat{T}_3+\alpha \hat{Y}/2)\bigl)\hat{V}_\pm \\
&[\hat{U}_\pm, \hat{V}_\mp]=\pm \bigl(\Phi_3(s_3, \hat{H}/3-\alpha \hat{Y}+\alpha)-\Phi_3(s_3, \hat{H}/3-\alpha \hat{Y})\bigl) \hat{T}_\mp \, ,
\label{polyend}
\end{align}
where we introduced the three functions: 
\begin{align}
\Xi_{12}^{(s_1, s_2)}(\hat{H},\hat{T}_3, \hat{Y})&:=\Phi_1(s_1, \hat{H}/3 +\alpha \hat{T}_3+\alpha \hat{Y}/2)\Phi_2(s_2, \hat{H}/3 -\alpha \hat{T}_3+\alpha \hat{Y}/2+\alpha)\\
\Xi_{13}^{(s_1, s_3)}(\hat{H},\hat{T}_3, \hat{Y})&:=\Phi_1(s_1, \hat{H}/3+\alpha \hat{T}_3+\alpha \hat{Y}/2)\Phi_3(s_3, \hat{H}/3-\alpha \hat{Y}+\alpha)\\
\Xi_{23}^{(s_2, s_3)}(\hat{H},\hat{T}_3, \hat{Y})&:=\Phi_2(s_2, \hat{H}/3-\alpha \hat{T}_3+\alpha\hat{Y}/2)\Phi_3(s_3, \hat{H}/3-\alpha \hat{Y}+\alpha) \, ,
\label{funs}
\end{align}
and $(\hat{T}_-)^\dagger =  \hat{T}_+ \, , (\hat{U}_-)^\dagger =  \hat{U}_+ \, , (\hat{V}_-)^\dagger =  \hat{V}_+$ whereas $\hat{H}^\dagger = \hat{H} \, , \, \hat{T}_3^\dagger = \hat{T}_3 \, , \hat{Y}^\dagger = \hat{Y}$. 

\noindent The specific choice \eqref{cartan} is dictated by the following reasoning. Consider the Hamiltonian of the isotropic harmonic oscillator \eqref{ho} with $n=3$, i.e. $s_1=s_2=s_3=\alpha=1$. In this limiting case,  the polynomial algebra collapses to:
\begin{equation}
[\hat{H},\cdot]=[\hat{T}_3,\hat{Y}]=0 
\label{cartanham}
\end{equation}
\begin{align}
[\hat{T}_{+},\hat{T}_-]&=2\hat{T}_3 \hskip 2cm
[\hat{T}_3,\hat{T}_\pm]=\pm\hat{T}_\pm \hskip 1.6cm [\hat{Y},\hat{T}_\pm]=0 \\
[\hat{V}_{+},\hat{V}_-]&=\hat{T}_3+\frac{3}{2}\hat{Y} \hskip 1.225cm [\hat{T}_3,\hat{V}_\pm]=\pm\frac{1}{2}\hat{V}_\pm \hskip 1.3cm [\hat{Y},\hat{V}_\pm]=\pm \hat{V}_\pm\\
 [\hat{U}_{+},\hat{U}_-]&=-\hat{T}_3+\frac{3}{2}\hat{Y}\hskip 0.95cm [\hat{T}_3,\hat{U}_\pm]=\mp\frac{1}{2}\hat{U}_\pm \hskip 1.2cm [\hat{Y},\hat{U}_\pm]=\pm \hat{U}_\pm \, \\ 
[\hat{T}_\pm, \hat{V}_\mp]&=\mp \hat{U}_\mp \hskip 1.75cm [\hat{T}_{\pm},\hat{U}_\pm]=\pm \hat{V}_\pm \hskip 1.3cm[\hat{U}_\pm, \hat{V}_\mp]=\pm\hat{T}_\mp \, .
\label{eq:su3}
\end{align} 
The commutation relations involving the generators $\{ \hat{T}_3,\hat{Y}, \hat{T}_\pm, \hat{U}_{\pm}, \hat{V}_\pm\}$ are those of the $\mathfrak{su}(3)$ Lie algebra \cite{greiner1994quantum}, ($\hat{T}_3,\hat{Y}$) being the Cartan generators, which is known to be the symmetry algebra of the three-dimensional isotropic harmonic oscillator \cite{doi:10.1119/1.1971373, Wybourne1974ClassicalGF}. In particular, the change of basis:
\begin{align}
\hat{X}_1&=\frac{1}{2}(\hat{T}_{+}+\hat{T}_{-}) \qquad \hat{X}_4=\frac{1}{2}\bigl(\hat{V}_{+}+\hat{V}_{-})\qquad
\hat{X}_6=\frac{1}{2}(\hat{U}_{+}+\hat{U}_{-})\qquad \hat{X}_3=\hat{T}_3\\
\hat{X}_2&=\frac{\imath}{2} (\hat{T}_{-}-\hat{T}_{+})\qquad
\hat{X}_5=\frac{\imath}{2}(\hat{V}_{-}-\hat{V}_{+})\qquad
\hat{X}_7=\frac{\imath}{2} (\hat{U}_{-}-\hat{U}_{+})\qquad
\hat{X}_8=\frac{\sqrt{3}}{2}\hat{Y} \, ,
\end{align}
allows us to recover the Hermitian generators $\hat{X}_a$ ($a=1, \dots, 8$) as:
\begin{equation}
\hat{X}_a=\frac{1}{2}\hat{A}^\dagger_i[\lambda_a]_{ij}\hat{A}_j \, ,
\label{eq:gens}
\end{equation}
where $\lambda_a$ are the eight Gell-mann matrices \cite{georgi1999lie}. In this basis, the commutation relations read:
\begin{equation}
[\hat{X}_a, \hat{X}_b]= \imath f_{abc} \hat{X}_c \, ,
\label{eq:gellman}
\end{equation}
with $f_{123}=1$, $f_{147}=f_{165}=f_{246}=f_{257}=f_{345}=f_{376}=1/2$, $f_{458}=f_{678}=\sqrt{3}/2$.
In the above realisation, the quantum integrals of motion arise as the following combinations:
\begin{align}
\hat{X}_1&=\hat{F}_{12} \qquad \quad \, \hat{X}_4=\hat{F}_{13}\qquad\quad
\,\,\,\hat{X}_6=\hat{F}_{23}\qquad \quad \,\,\hat{X}_3=(\hat{F}_{11}-\hat{F}_{22})/2\\
\hat{X}_2&=\hat{L}_{12}/2\qquad
\,\,\hat{X}_5=\hat{L}_{13}/2\qquad
\,\,\hat{X}_7=\hat{L}_{23}/2\qquad
\,\,\hat{X}_8=(\hat{F}_{11}+\hat{F}_{22}-2\hat{F}_{33})/2\sqrt{3} \, ,
\end{align}
\noindent where $\hat{F}_{ij}:=\frac{1}{2}(\hat{p}_i \hat{p}_j+\hat{x}_i\hat{x}_j)$ is the Demkov-Fradkin tensor \cite{ doi:10.1119/1.1971373, Dem} and $\boldsymbol{\hat{L}}=(\hat{L}_{23}, -\hat{L}_{13}, \hat{L}_{12})$ with $\hat{L}_{ij}=\hat{x}_i \hat{p}_j-\hat{x}_j \hat{p}_i$,  is the angular momentum vector. In this basis, the  second-order and third-order Casimir invariants read:
\begin{align}
 \hat{C}^{(2)}_{\mathfrak{su}(3)}&= \hat{X}_1^2+ \hat{X}_2^2+ \hat{X}_3^2+ \hat{X}_4^2+ \hat{X}_5^2+ \hat{X}_6^2+ \hat{X}_7^2+ \hat{X}_8^2 \\
\hat{C}_{\mathfrak{su}(3)}^{(3)}&=\frac{3}{2}\hat{X}_3-\frac{\sqrt{3}}{2}\hat{X}_8+\sqrt{3} (\hat{X}_1^2\hat{X}_8+ \hat{X}_2^2  \hat{X}_8+\hat{X}_3^2  \hat{X}_8)+3 (\hat{X}_1 \hat{X}_4  \hat{X}_6+  \hat{X}_1  \hat{X}_5  \hat{X}_7-  \hat{X}_2  \hat{X}_4 \hat{X}_7+  \hat{X}_2  \hat{X}_5  \hat{X}_6) \nonumber \\
&\,+\frac{3}{2}  (\hat{X}_3  \hat{X}_4^2+ \hat{X}_3 \hat{X}_5^2- \hat{X}_3  \hat{X}_6^2-  \hat{X}_3  \hat{X}_7^2)-\frac{\sqrt{3}}{2}(\hat{X}_4^2  \hat{X}_8+ \hat{X}_5^2 \hat{X}_8+\hat{X}_6^2 \hat{X}_8+ \hat{X}_7^2 \hat{X}_8)-\frac{1}{\sqrt{3}} \hat{X}_8^3 \, ,
\end{align}
and result in the following expressions involving the Hamiltonian operator:
\begin{align}
\hat{C}^{(2)}_{\mathfrak{su}(3)}=\frac{1}{3}(\hat{H}^2-9/4) \, , \qquad 
\hat{C}_{\mathfrak{su}(3)}^{(3)}= \frac{1}{9}(\hat{H}^2-9/4)\hat{H} \, .
\end{align}
The existence of Casimir invariants for the general polynomial algebra defined above is an open problem. 
\begin{rem}
If we consider Hamiltonian systems such as for $m_1=m_2=m_3=1$ they describe the standard harmonic oscillator, then the polynomial algebra we have constructed can be considered as a polynomial deformation of the  Lie algebra $\mathfrak{u}(1) \oplus \mathfrak{su}(3)$ involving higher order powers of both the Cartan generators and the Hamiltonian, the latter being a central element.  Let us consider, as an example, the three-dimensional Hamiltonian system:
\begin{equation}
\hat{H}=\hat{H}_1+\hat{H}_2+\hat{H}_3=\frac{1}{2} \bigl(-\partial_{x_1x_1}+x^2_1/m_1^2\bigl)+\frac{1}{2} \bigl(-\partial_{x_2x_2}+x^2_2/m_2^2\bigl)+\frac{1}{2} \bigl(-\partial_{x_3x_3}+x^3_3/m_3^2\bigl) \, ,
\label{eq:hamani}
\end{equation}
where $m_1, m_2, m_3$ are mutually prime natural numbers. The ladder operators are first-order:
\begin{equation}
\hat{A}_i=\frac{\partial_{x_i}+x_i/m_i }{\sqrt{2}} \, , \qquad \hat{A}^\dagger_i=\frac{-\partial_{x_i}+x_i/m_i }{\sqrt{2}}  \, ,
\label{anisotropic}
\end{equation}
and satisfy the commutation relations:
\begin{align}
[\hat{H}_i, \hat{A}_i]&=-\frac{1}{m_i} \hat{A}_i \hskip 1.65cm [\hat{H}_i, \hat{A}^{\dagger}_i]= \frac{1}{m_i}\hat{A}^\dagger_i \label{eq:ladderaniso1} \\
\hat{A}_i^{\dagger}\hat{A}_i&=\hat{H}_i-1/(2m_i) \qquad \quad \,\,\hat{A}_i\hat{A}_i^\dagger\!=\hat{H}_i+1/(2m_i)  \, .
\label{eq:ladderaniso2}
\end{align}
\noindent Since  $\alpha_k=1/m_k$, the condition $\alpha_i s_i=\alpha_j s_j=\alpha$ is satisfied for $s_k=m_k$, $k=1,2,3$ and $\alpha=1$. This means that the generators of the polynomial algebra  \eqref{poly1}-\eqref{polyend} take the form:
\begin{equation}
\hat{H}=\hat{H}_1+\hat{H}_2+\hat{H}_3 \, ,\quad \hat{T}_3=\frac{\hat{H}_1-\hat{H}_2}{2} \, , \quad \hat{Y}=\frac{\hat{H}_1+\hat{H}_2-2\hat{H}_3}{3} \, , \quad \hat{E}_{ij}=(\hat{A}^\dagger_i)^{m_i} (\hat{A}_j)^{m_j} \quad (i \neq j) \, .
\label{eq:generatorspoly}
\end{equation}
\noindent In this case, the structure functions are given by:  
\begin{align}
\Xi_{12}^{(m_1, m_2)}(\hat{H},\hat{T}_3, \hat{Y})&:=\Phi_1(m_1, \hat{H}/3 +\hat{T}_3+ \hat{Y}/2)\Phi_2(m_2, \hat{H}/3 - \hat{T}_3+ \hat{Y}/2+1)\\
\Xi_{13}^{(m_1, m_3)}(\hat{H},\hat{T}_3, \hat{Y})&:=\Phi_1(m_1, \hat{H}/3+\hat{T}_3+ \hat{Y}/2)\Phi_3(m_3, \hat{H}/3- \hat{Y}+1)\\
\Xi_{23}^{(m_2, m_3)}(\hat{H},\hat{T}_3, \hat{Y})&:=\Phi_2(m_2, \hat{H}/3- \hat{T}_3+\hat{Y}/2)\Phi_3(m_3, \hat{H}/3-\hat{Y}+1) \, ,
\label{funsanis}
\end{align}
\noindent with:
 \begin{align}
\Phi_1(m_1, \hat{H}/3 +\hat{T}_3+ \hat{Y}/2)&=\prod_{i=1}^{m_1} \biggl(\hat{H}/3 +\hat{T}_3+ \hat{Y}/2-\frac{2i-1}{2m_1}\biggl) \\
\Phi_2(m_2, \hat{H}/3 -\hat{T}_3+ \hat{Y}/2)&=\prod_{i=1}^{m_2} \biggl(\hat{H}/3 -\hat{T}_3+ \hat{Y}/2-\frac{2i-1}{2m_2}\biggl) \\
\Phi_3(m_3, \hat{H}/3 - \hat{Y})&=\prod_{i=1}^{m_3} \biggl(\hat{H}/3 -\hat{Y}-\frac{2i-1}{2m_3}\biggl) \, .
\end{align}

\noindent For $m_1=m_2=m_3=1$ the three-dimensional isotropic harmonic oscillator is recovered.

\end{rem}

\subsection{Rational extension of the harmonic oscillator associated with type III Hermite exceptional orthogonal polynomials}
\label{ssec3.1}
 
 \noindent   What we are interested to investigate in this Section \ref{ssec3.1} is a $3$D rational extension of the harmonic oscillator associated with Hermite exceptional orthogonal polynomials (EOP) of type III \cite{doi:10.1063/1.4798807, doi:10.1063/1.3671966, GomezUllate2004, Carinena_2008, Fellows_2009,  doi:10.1063/1.3651222, GRANDATI20122411}.  The two-dimensional version of this problem has been investigated in \cite{doi:10.1063/1.4823771}, where a new pair of ladder operators for the one-dimensional rationally extended harmonic oscillator have been introduced.  Here, we deal with the three-dimensional case and, in particular, we consider the Hamiltonian system in \eqref{ham3D} with:
 \begin{equation}
 \hat{H}_1 =-\partial_{x_1 x_1}+x_1^2-2\biggl[\frac{\mathcal{H}''_m}{\mathcal{H}_m}-\biggl(\frac{\mathcal{H}'_m}{\mathcal{H}_m}\biggl)^2 + \,1 \biggl] \,  , \quad \hat{H}_2 =-\partial_{x_2 x_2}+x_2^2 \, , \quad \hat{H}_3 =-\partial_{x_3 x_3}+x_3^2
 \label{eq:rationallyextended}
 \end{equation}
 where $\mathcal{H}_m:=(-\imath)^m H_m(\imath x)$,  $H_m(z)$ being the usual Hermite orthogonal polynomials with $m$ even.
The one-dimensional constituents of this three-dimensional Hamiltonian system are two standard harmonic oscillators in the $x_2$ and $x_3$ axes and a rationally extended oscillator in the $x_1$ axis. In one dimension, the latter arises as a superpartner of the standard Hermite harmonic oscillator in the framework of Supersymmetric Quantum Mechanics (SUSYQM) \cite{Bagchi, CooperAvidane}. Let us consider the following pair of SUSY partners:
  \begin{equation}
 \hat{H}_+ := -\frac{\text{d}^2}{\text{d}x^2}+V_+(x)-E_m \, , \quad  \hat{H}_- := -\frac{\text{d}^2}{\text{d}x^2}+V_-(x)-E_m \, ,
 \label{eq:susypartner}
 \end{equation}
 where the two potentials $V_\pm(x) := W^2(x)\mp W'(x)+E_m$ are given in terms of the superpotential  $W(x)=-\phi'_m(x)/\phi_m(x)$.  Here $E_m:=-2m-1$, $\phi_m (x)=\mathcal{H}_m(x)\exp(x^2/2)$ is solution of the equation $\hat{H}_+ \phi_m(x)=0$ and the partner potentials read:
\begin{equation} 
 V_+(x)=x^2 \, , \quad
 V_-(x)=x^2 -2\biggl(\frac{\mathcal{H}_m''}{\mathcal{H}_m}-\biggl(\frac{\mathcal{H}_m'}{\mathcal{H}_m}\biggl)^2\,+\,1\biggl) \, .
 \end{equation}
\noindent  These two Hamiltonians are connected through the action of the first-order ladder operators:
  \begin{equation}
 a:=\frac{\text{d}}{\text{d}x}+W(x)=\frac{\text{d}}{\text{d}x}-\biggl(x+\frac{\mathcal{H}'_m}{\mathcal{H}_m} \biggl) \, , \quad a^{\dagger}=-\frac{\text{d}}{\text{d}x}+W(x)=-\frac{\text{d}}{\text{d}x}-\biggl(x+\frac{\mathcal{H}'_m}{\mathcal{H}_m} \biggl)  \, ,
 \label{eq:ladderope}
 \end{equation}
 as they interwine as:
 \begin{equation}
 \hat{a} \hat{H}_+=\hat{H}_- \hat{a} \, , \quad \hat{a}^\dagger \hat{H}_-=\hat{H}_+ \hat{a}^\dagger \, .
 \label{eq:interw}
 \end{equation}
 
 \noindent In particular, for even values of $m$, the eigenfunctions and eigenvalues of the two superpartner $\hat{H}_\pm$ turn out to be:
\begin{equation}
\begin{cases}
\psi_n^+(x)=\mathcal{N}_n^+ H_n(x)\exp(-x^2/2) \, , \qquad \hskip 0.15cm E_{n}^+=2 (n+m+1) \qquad (n = 0, 1, 2, \dots) \\
\psi_n^-(x)=\mathcal{N}_n^- \frac{y_{n+m+1}^{(m)}}{\mathcal{H}_m(x)} \exp(-x^2/2) \, , \qquad E_{n}^-=2 (n+m+1) \qquad (n = -m-1, 0, 1, 2, \dots) 
\end{cases}
\label{eigen}
\end{equation}
where  $y_0^{(m)}(x)=1$ and $y_{k+m+1}^{(m)}(x)=-\mathcal{H}_m(x)H_{k+1}(x)-2m \mathcal{H}_{m-1}(x)H_{k}(x) \quad (k = 0,1, 2, \dots)$ are given in terms of the type III Hermite EOP $y_{\mathsf{n}}^{(m)}(x)$, with $\mathsf{n}:=n+m+1$. The normalisation constants read:
\begin{align*}
\mathcal{N}_{n}^+:=(\sqrt{\pi}2^{n} n!)^{-1/2} \quad \text{and}\quad 
&\mathcal{N}_n^-:=
\begin{cases}
(\frac{2^m m!}{\sqrt{\pi}})^{1/2}  & \quad \text{if } n=-m-1 \\
(\sqrt{\pi}2^{n+1}(n+m+1)n!)^{-1/2}  & \quad \text{if } n=0,1,2,\dots \, .
\end{cases}
\end{align*}
\noindent This model is characterized by an additional bound state placed below the oscillator spectrum and associated to the energy eigenvalue $E_{-m-1}=-2m-1$. 
The eigenfunctions defined above are orthonormal in $L^2(\mathbb{R})$ w.r.t. the positive definite measure $\text{d}\mu(x):=\exp(-x^2)/(\mathcal{H}_m)^2 \text{d}x$. In the work \cite{doi:10.1063/1.4823771}, the authors showed the existence of a pair of lowering and raising operators ($\hat{A}_m$, $\hat{A}_m^\dagger$) for the Hamiltonian $\hat{H}_-$ of the rationally extended oscillator. These ladder operators are constructed through the composite action of the first-order ladder operators \eqref{eq:ladderope} and the additional $m$ first-order auxiliary operators:
\begin{equation}
a_i:=\frac{\text{d}}{\text{d}x}+\biggl(x+\frac{\mathcal{H}'_{i-1}}{\mathcal{H}_{i-1}}- \frac{\mathcal{H}'_{i}}{\mathcal{H}_{i}} \biggl) \, , \quad a_i^{\dagger}=-\frac{\text{d}}{\text{d}x}+\biggl(x+\frac{\mathcal{H}'_{i-1}}{\mathcal{H}_{i-1}}- \frac{\mathcal{H}'_{i}}{\mathcal{H}_{i}} \biggl) \qquad (i=1, \dots, m) \, ,
\label{supercharges}
\end{equation}
such as:
\begin{equation}
\hat{a}_i^\dagger \hat{a}_i= \hat{H}_i \, , \quad \hat{a}_i \hat{a}_i^\dagger= \hat{H}_{i+1}+2 \, , \quad \text{with}\quad \hat{H}_i =-\frac{\text{d}^2}{\text{d}x^2} +x^2-2\biggl(\frac{\mathcal{H}_{i-1}''}{\mathcal{H}_{i-1}}-\biggl(\frac{\mathcal{H}'_{i-1}}{\mathcal{H}_{i-1}}\biggl)^2	\,\biggl) -	\,3  \, .
\label{eq:intcharges}
\end{equation}
Specifically, by considering the recursive application of the $m+1$ first-order operators \eqref{eq:ladderope} and \eqref{supercharges}, it is possible to define the higher order lowering and raising operators:
\begin{equation}
\hat{A}_m:= \biggl(\prod_{i=1}^m \hat{a}_{m-i+1}\biggl) \hat{a}^\dagger \qquad  \hat{A}^\dagger_m:= \hat{a} \biggl(\prod_{i=1}^m \hat{a}^\dagger_{i}\biggl)  \, ,
\label{eq:prod}
\end{equation}
\noindent for which hold:
\begin{equation}
[\hat{H}_-, \hat{A}_{m}]=-(2m+2) \hat{A}_{m}\,  \quad [\hat{H}_-, \hat{A}^\dagger_{m}]=(2m+2) \hat{A}^\dagger_{m} \,  \quad \, [\hat{A}_{m}, \hat{A}^\dagger_{m}]=\phi(\hat{H}_-+2m+2)-\phi(\hat{H}_-) \, ,
\label{eq:high-order}
\end{equation}
where the following polynomial in $\hat{H}_-$ has been introduced:
\begin{equation}
\mathcal{\phi}(\hat{H}_-):=(\hat{H}_-)\prod_{i=1}^{m}(\hat{H}_--2m-2-2i) \, .
\label{eq:polym} 
\end{equation}
In \eqref{eq:prod} we used the equality $\bigl(\prod_{i=1}^m \hat{a}_{m-i+1}\bigl)^\dagger = \prod_{i=1}^m \hat{a}^\dagger_{i}$. Then, this new pair of $(m+1)$th-order ladder operators, together with $\hat{H}_-$, closes in a polynomial Heisenberg algebra of order $m$. These are the ladder operators we will be interested in this paper in relation to our polynomial algebra. We will specify later  in the Section their explicit action on the eigenstates of the rationally extended oscillator.

\noindent The 3D Hamiltonian that we have introduced arises in terms of the one-dimensional superpartner Hamiltonians:
\begin{equation}
\begin{cases}
\hat{H}_1 := \hat{H}_--2m-1 \quad \text{in the} \,\, x_1-\text{axis} \quad (\text{rationally extended oscillator}) \\
\hat{H}_2 := \hat{H}_+-2m-1 \quad \text{in the} \,\, x_2-\text{axis} \quad (\text{standard oscillator})\\
\hat{H}_3 := \hat{H}_+-2m-1 \quad \text{in the} \,\, x_3-\text{axis} \quad (\text{standard oscillator}) \, .
\label{eq:3d}
\end{cases}
\end{equation}

\noindent Clearly, due to Cartesian separability,  the eigenfunctions and eigenvalues associated to the three-dimensional Hamiltonian operator $\hat{H}=\hat{H}_1+\hat{H}_2+\hat{H}_3$ turns out to be:
\begin{align}
\Psi_{n_1, n_2, n_3}^{(m)}(x_1, x_2, x_3)&=\mathcal{N}_{n_1} \mathcal{N}_{n_2} \mathcal{N}_{n_3} \frac{y^{(m)}_{n_1+m+1}(x_1)}{\mathcal{H}_m(x_1)}H_{n_2}(x_2)H_{n_3}(x_3)e^{-\frac{1}{2}(x_1^2+x_2^2+x_3^2)} \\
E_{n_1, n_2, n_3}&=2(n_1+n_2+n_3)+3
\label{eq:eigen}
\end{align}
with $n_1=-m-1, 0, 1,2, \dots \, , n_2=0,1,2,\dots \, , n_3=0,1,2,\dots \, \,$. 
Taking into account our previous discussion, let us introduce the following pairs of ladder operators in the three directions $(x_1,x_2,x_3)$:
\begin{align}
\hat{A}_{1;m}&=\hat{a}_m \dots \hat{a}_1 \hat{a}^\dagger \,  \qquad \quad \,\, \hat{A}_{1;m}^{\dagger}=\hat{a}	\, \hat{a}_1^\dagger \dots \hat{a}_m^\dagger  \\
\hat{A}_2&=\partial_{x_2}+x_2 \,  \hskip 1.9cm \hat{A}_2^\dagger=-\partial_{x_2}+x_2\\
\hat{A}_3&=\partial_{x_3}+x_3 \,  \hskip 1.9cm \hat{A}_3^\dagger=-\partial_{x_3}+x_3 \, ,
\end{align}
\noindent satisfying, together with their corresponding one-dimensional Hamiltonians, the commutation relations:
\begin{align}
[\hat{H}_1, \hat{A}_{1;m}]&=-(2m+2) \hat{A}_{1;m}\,  \quad [\hat{H}_1, \hat{A}^\dagger_{1;m}]=(2m+2) \hat{A}^\dagger_{1;m} \,  \quad \, [\hat{A}_{1;m}, \hat{A}^\dagger_{1;m}]=\phi_1(\hat{H}_1+2m+2)-\phi_1(\hat{H}_1)\\
[\hat{H}_2, \hat{A}_{2}]&=-2 \hat{A}_{2}\,  \hskip 2.2cm [\hat{H}_2, \hat{A}^\dagger_{2}]=2 \hat{A}^\dagger_{2} \,  \hskip 2.575cm [\hat{A}_{2}, \hat{A}^\dagger_{2}]=2\\
[\hat{H}_3, \hat{A}_{3}]&=-2 \hat{A}_{3}\,  \hskip 2.3cm[\hat{H}_3, \hat{a}^\dagger_{3}]=2 \hat{A}^\dagger_{3} \, \hskip 2.575cm [\hat{A}_{3}, \hat{A}^\dagger_{3}]=2 \, ,
\end{align}
where: \begin{equation}
\phi_1(\hat{H}_1)=(\hat{H}_1+2m +1)\prod_{i=1}^{m}(\hat{H}_1-1-2i)
\label{eq:poly} \, .
\end{equation}
Given the above relations, we can see that the equalities $2(m+1) s_1 = 2 s_2 = 2 s_3 = \alpha$ are satisfied for the values $s_1=1, s_2=s_3=m+1$ and $\alpha = 2(m+1)$. Thus, from the general results, we obtain that the following operators:
\begin{align}
&\hat{T}_3:=\frac{\hat{H}_1-\hat{H}_2}{4(m+1)} \qquad \qquad \quad \hat{Y}:=\frac{\hat{H}_1+\hat{H}_2-2\hat{H}_3}{6(m+1)} \, ,
\end{align}
together with the higher order ones:
\begin{align}
& \hat{T}_{+}:=\hat{A}^\dagger_{1;m} (\hat{A}_2)^{m+1}  \qquad  \hat{U}_{+}:=(\hat{A}^\dagger_{2})^{m+1} (\hat{A}_3)^{m+1}  \qquad \hat{V}_{+}:=\hat{A}^\dagger_{1;m} (\hat{A}_3)^{m+1}   \\
&\hat{T}_{-}:= (\hat{A}^\dagger_2)^{m+1}\hat{A}_{1;m}  \qquad\hat{U}_{-}:= (\hat{A}^\dagger_{3})^{m+1} (\hat{A}_2)^{m+1} \qquad \hat{V}_{-}:= (\hat{A}^\dagger_3)^{m+1} \hat{A}_{1;m} \, ,
\label{eq:com}
\end{align}
\noindent commute with $\hat{H}$ and close to give the polynomial algebra \eqref{polcar}-\eqref{funs} with structure functions given by:
\begin{align*}
 \Xi_{12}^{(1, m+1)}(\hat{H},\hat{T}_3, \hat{Y})&=\Phi_1(1, \hat{H}/3 +2(m+1)\hat{T}_3+(m+1) \hat{Y})\Phi_2(m+1, \hat{H}/3 -2(m+1) \hat{T}_3+ (m+1)\hat{Y}+2(m+1))\\
\Xi_{13}^{(1, m+1)}(\hat{H},\hat{T}_3, \hat{Y})&=\Phi_1(1, \hat{H}/3+2(m+1)\hat{T}_3+ (m+1)\hat{Y})\Phi_3(m+1, \hat{H}/3- 2(m+1)\hat{Y}+2(m+1))\\
\Xi_{23}^{(m+1, m+1)}(\hat{H},\hat{T}_3, \hat{Y})&=\Phi_2(m+1, \hat{H}/3- 2(m+1)\hat{T}_3+(m+1)\hat{Y})\Phi_3(m+1, \hat{H}/3-2(m+1)\hat{Y}+2(m+1)) \, ,
\label{funsext}
\end{align*}
\noindent where:
\begin{align}
\Phi_1(1, \hat{H}/3 +2(m+1)\hat{T}_3+ (m+1)\hat{Y})&=(\hat{H}/3 +2(m+1)\hat{T}_3+ (m+1)\hat{Y}+2m +1) \nonumber \\& \hskip 0.8cm  \times \prod_{i=1}^{m}(\hat{H}/3 +2(m+1)\hat{T}_3+ (m+1)\hat{Y}-1-2i) \\
\Phi_2(m+1, \hat{H}/3 -2(m+1)\hat{T}_3+ (m+1)\hat{Y}) & =\prod_{i=1}^{m+1} \biggl(\hat{H}/3 -2(m+1)\hat{T}_3+ (m+1)\hat{Y}-2i+1\biggl) \\
\Phi_3(m+1, \hat{H}/3 - 2(m+1)\hat{Y}) & =\prod_{i=1}^{m+1} \biggl(\hat{H}/3 -2(m+1)\hat{Y}-2i+1\biggl) \, .
\end{align}
\subsubsection{Degeneracies of energy levels and  irreps of the polynomial algebra}
\label{sssec}

\noindent The construction of finite-dimensional representations for the polynomial algebras is in general a difficult task. It represents however an important problem as the finite-dimensional representations, besides providing the degeneracies,  can also offer constraints on the spectrum of the models for which other methods are not in general applicable, in particular when the potential involves special functions defined only by non-linear ODEs. In such cases, algebraic approaches offer powerful tools for providing insights into the spectrum. Then, there is a need of developing further algebraic methods.

The construction of finite-dimensional representations of polynomial algebras have been so far limited mostly to the context of two-dimensional quantum superintegrable systems and their related rank-$1$ polynomial algebras with three generators. For such cases the approach due to Daskaloyannis has been applied widely. It consists of constructing the corresponding deformed oscillator algebras \cite{Bonatsos:1994fi,PhysRevA.50.3700, BONATSOS1999537, doi:10.1063/1.1348026, Bonatsos:1993st, 2011SIGMA...7..054T}.



It was pointed out in \cite{doi:10.1063/1.4823771} that the analysis of solutions by the Daskaloyannis approach can be complicated because it requires solving the various algebraic constraints and removing non physical solutions. In \cite{Marquette2014} It was shown that a more direct approach of constructing finite-dimensional representations by using zero modes of integrals allow one to treat systems with higher order integrals and related algebras more efficiently. This is the approach we plan to adapt and develop in this Section. The higher rank polynomial algebras we have introduced display interesting properties that will make the construction of representations feasible. In Section \ref{sec3}, we have been able to identify integrals playing the role of Cartan, raising and lowering operators, respectively. We will also develop ideas which have similarity with the induced module construction. This will allow us to describe non-trivial degeneracies in terms of states connected by the higher order quantum integrals, the latter being referred to as multiplets. The actions of these higher order operators on eigenstates can be found by combining the specific actions of their constituent ladder operators, which reads:
\begin{align*}
\hat{A}^\dagger_{1;m} \ket{n_1}_{x_1}&=
\begin{cases} 
\sqrt{2^{m+1} m!} \ket{0}_{x_1} \hskip 7cm\text{if} \quad n_1 = -m-1\\
-\sqrt{2^{m+1} (n_1+2m+2)\prod_{i=0}^{m-1}(n_1+m-i)}  \ket{n_1+m+1}_{x_1} \hskip 0.4cm \text{if} \quad n_1 = 0, 1, 2, \dots
\end{cases} \\
\hat{A}_{1;m} \ket{n_1}_{x_1}&=
\begin{cases} 
0 \hskip 9.22cm \text{if} \quad n_1 =-m-1, 1,2\dots, m\\
\sqrt{2^{m+1} m!} \ket{-m-1}_{x_1} \hskip 6.05cm\text{if} \quad n_1 =0\\
-\sqrt{2^{m+1} (n_1+m+1)\prod_{i=0}^{m-1}(n_1-i-1)}  \ket{n_1-m-1}_{x_1} \hskip 0.8cm \text{if} \quad n_1 = m+1,m+2,\dots
\end{cases} \\
(\hat{A}_2^\dagger)^{m+1} \ket{n_2}_{x_2}&=
\sqrt{\prod_{j=1}^{m+1}(n_2+j)}  \ket{n_2+m+1}_{x_2} \hskip 5.15cm \text{if} \quad n_2 = 0, 1, 2,  \dots\\
(\hat{A}_2)^{m+1} \ket{n_2}_{x_2}&=
\begin{cases} 
0 \hskip 9.3cm\text{if} \quad n_2 = 0,1,2, \dots, m\\
\sqrt{\prod_{j=0}^{m}(n_2-j)}  \ket{n_2-m-1}_{x_2} \hskip 4.7cm \text{if} \quad n_2 = m+1, m+2,  \dots
\end{cases} \\
(\hat{A}_3^\dagger)^{m+1} \ket{n_3}_{x_3}&=
\sqrt{\prod_{j=1}^{m+1}(n_3+j)}  \ket{n_3+m+1}_{x_3} \hskip 5.2cm \text{if} \quad n_3 = 0, 1, 2,  \dots\\
(\hat{A}_3)^{m+1} \ket{n_3}_{x_3}&=
\begin{cases} 
0 \hskip 9.35cm\text{if} \quad n_3 = 0,1,2, \dots, m\\
\sqrt{\prod_{j=0}^{m}(n_3-j)}  \ket{n_3-m-1}_{x_3} \hskip 4.75cm \text{if} \quad n_3 = m+1, m+2,  \dots
\end{cases} 
\end{align*}
\noindent Observing the action of the ladder operators in the $x_1$ axis, and considering the standard action the ladder have in the other two axes, we see that the (unnormalized) basis states are given by:

\begin{align*}
\ket{n_1}_{x_1}\ket{n_2}_{x_2}\ket{n_3}_{x_3}=\begin{cases}
 [(\hat{A}_{1;m}^\dagger)^k\ket{m-1}_{x_1}] \otimes [(\hat{A}_2^\dagger)^{n_2}\ket{0}_{x_2}] \otimes [(\hat{A}_3^\dagger)^{n_3} \ket{0}_{x_3}] \,\propto\,\! \ket{(m+1)(k-1)}_{x_1}\ket{n_2}_{x_2}\ket{n_3}_{x_3} \\
 [(\hat{A}_{1;m}^\dagger)^k\ket{1}_{x_1}] \otimes [(\hat{A}_2^\dagger)^{n_2}\ket{0}_{x_2}] \otimes [(\hat{A}_3^\dagger)^{n_3} \ket{0}_{x_3}] \,\,\qquad\propto \ket{1+k(m+1)}_{x_1}\ket{n_2}_{x_2}\ket{n_3}_{x_3}  \\
\vdots \hskip 7.8cm \vdots\\
[(\hat{A}_{1;m}^\dagger)^k\ket{m}_{x_1}] \otimes [(\hat{A}_2^\dagger)^{n_2}\ket{0}_{x_2}] \otimes [(\hat{A}_3^\dagger)^{n_3} \ket{0}_{x_3}] \qquad \propto \ket{m+k(m+1)}_{x_1}\ket{n_2}_{x_2}\ket{n_3}_{x_3} \, ,
\end{cases}
\label{eq:bstates}
\end{align*}
for $k=0,1\dots$, $n_2=0,1,\dots$ and $n_3=0,1,\dots\,\,.$

\begin{rem}
	This decomposition is due to the fact that we are using this specific type of ladder operators in the $x_1$ axis, for which it is known the polynomial Heisenberg algebra possesses $m + 1$ infinite-dimensional unirreps spanned by the states $\ket{j+(m+1)k}_{x_1}$, for $j=-m-1, 1, \dots, m$ and $k=0,1,\dots\,\,$. Other types of ladder operators exist in the literature for the rationally extended Hermite oscillator associated with type III Hermite EOP \cite{ doi:10.1063/1.1853203,2011_SQM, carplyu, carplyu2}, for example, the pair \cite{doi:10.1063/1.4798807}:
	\begin{equation}
	\hat{B}_{m} = \hat{a} \, \hat{b} \, \hat{a}^\dagger \, \quad \hat{B}^{\dagger}_{m}=\hat{a} \, \hat{b}^\dagger \,\hat{a}^\dagger \, ,
	\label{ladders}
	\end{equation} 
	with $(\hat{a},\hat{a}^\dagger)$ given in \eqref{eq:ladderope} and ${b}=\frac{\text{d}}{\text{d}x}+x$, ${b}^\dagger=-\frac{\text{d}}{\text{d}x}+x$. However, as explained in \cite{Quesne2015}, the choice of higher order ladder operators \eqref{eq:prod} is the suitable one in the study of higher order superintegrable models that are characterized by one-dimensional potentials related to type III Hermite EOP.
\end{rem}
\begin{rem}
	The set composed by the nine operators is not of minimal order for this model. There are in fact two standard oscillators in the $x_2$ and $x_3$ axes and therefore one generator of rotations is conserved. This would lead to a pair of shift operators constructed with the product $\hat{A}_2^\dagger \hat{A}_3$ and Hermitian conjugate.
\end{rem}
\noindent If we set $N = n_1 + n_2+ n_3$, a direct computation shows that the degeneracy of the energy level $E_N=2N+3$ is:
\begin{equation}
\text{deg}(E_N)=\begin{cases}
1 \hskip 4cm \text{if} \quad N = -m-1 \\
2\hskip 4cm\text{if} \quad N = -m \\
\vdots \hskip 4.1cm \vdots\\
m+1\hskip 3.3cm \text{if}\quad N=-1\\
(N+2)(N+3)/2+m \hskip 0.75 cm \text{if}\quad N = 0, 1 , 2, \dots \,\, .  \\
\end{cases}
\label{eq:energydeg}
\end{equation}

\noindent At a fixed $N$, the  basis states $\ket{n_1}_{x_1} \ket{n_2}_{x_2} \ket{n_3}_{x_3}$ are connected by the action of the operators $(\hat{T}_\pm, \hat{U}_\pm, \hat{V}_\pm)$ and find a useful graphical representation in the plane $(t_3, y)$, the latter being the eigenvalues associated to $\hat{T}_3$ and $\hat{Y}$ operators in the above Cartesian basis, or equivalently, in the basis: $\ket{N, n_1, n_2}=\ket{n_1}_{x_1}\ket{n_2}_{x_2}\ket{N-(n_1+n_2)}_{x_3}$.

  In order to characterize degeneracies of the model in terms of finite-dimensional irreps of the polynomial algebra, we denote the eigenstates of $\hat{H}$ as $\ket{N, n_1, n_2}$ and act on these physical states with shift operators $\{\hat{T}_\pm, \hat{U}_\pm, \hat{V}_\pm\}$ to see how they are connected for each $N$, i.e. for each energy level. Following up on the approach developed in \cite{Marquette2014}, we try to characterize some specific irreps of the polynomial algebra by two integer numbers $(N, \mathbf{p})$ and their basis states as $\ket{N, \tau, \mathbf{p}; (\tilde{t}_3, \tilde{y})}$ where $\mathbf{p}:=(p+1)(p+2)/2$ specify the number of connected states, the pair  $(\tilde{t}_3, \tilde{y})$ represents the eigenvalues of the operators $\hat{T}_3+\alpha_3$ and  $\hat{T}_8+\alpha_8$, where $(\alpha_3, \alpha_8)$ are two representation-dependent constants and $\tau$ distinguishes among repeated representations specified by the same $\mathbf{p}$ value. Sets of connected states will be called \textquotedblleft multiplets\textquotedblright and, as we shall see in the following, they will be of triangular shape when represented in the $(t_3, y)$ plane. In particular, for each $N$ we require the existence of a vector $\ket{N,\tau, \mathbf{p}; (\tilde{t}_3^{max}, \tilde{y})}$ such as:
  \begin{equation}
  \hat{T}_+ \ket{N, \tau, \mathbf{p}; (\tilde{t}_3^{max}, \tilde{y})} = \hat{V}_+ \ket{N,\tau,\mathbf{p} ; (\tilde{t}_3^{max}, \tilde{y})} = 0 \, ,
  \label{highest}
  \end{equation}
  
  \noindent and suppose that after $p+1$ applications of the operator $\hat{V}_-$ we get:
  
  \begin{equation}
  (\hat{V}_-)^{p+1} \ket{N, \tau, \mathbf{p} ; (\tilde{t}_3^{max}, \tilde{y})} = 0 \, .
  \label{p}
  \end{equation}
  
 \noindent  This operation completely specify the integer $p$ and, consequently, the boundary of the multiplet $\mathbf{p}$.
 Then, to characterize all connected states inside a given multiplet we act with:
 \begin{equation}
(\hat{T}_-)^{a_2}(\hat{V}_-)^{a_1}   \ket{N, \tau, \mathbf{p}; (\tilde{t}_3^{max}, \tilde{y})}  \, ,
 \label{eq:states}
 \end{equation}
 for  $0 \leq a_1 \leq p$ and $0 \leq a_2 \leq p-a_1$.
  To induce these irreps we rely on the physical basis $\ket{N, n_1, n_2}$ and on the explicit action of the higher order integrals of motion on this basis. Specifically, for even $m$, we adopt the following reparametrization:
\begin{equation}
 N=n_1+n_2+n_3=(m+1)\lambda +\mu\equiv N(m;\lambda,\mu) \, ,
 \label{occnumber}
 \end{equation}
 
 \noindent for some appropriate values of $\lambda=-1,0,1,\dots$ and $\mu=0,1,\dots,m$. This is helpful as in this way we are able to split the analysis of degeneracies into four different regions of the energy spectrum specified by given values of $\lambda$ and $\mu$. This turn out to be useful to simplify the analysis of this particular problem.  The energy spectrum reads:
 \begin{equation}
 E_N|_{N=N(m;\lambda,\mu)}=2 N(m;\lambda,\mu)+3=2((m+1)\lambda+\mu)+3 \, ,
 \label{en}
 \end{equation}
 and it has to be intended, for fixed values of $\lambda=\lambda^*$ and $\mu=\mu^*$, just as a linear function of the even integer $m$. 
 \begin{table}[h!]
 	\begin{small}
 		\begin{center}
 			\begin{tabular}{|c|c |c |c |c|} 
 				\hline
 				&$\lambda$ &$\mu$ &$N$&$E_{N}$  \\ [0.1ex] 
 				\hline
 				\textbf{I}&$-1$&$0,1,\dots,m$& $-m-1,-m,-m+1,\dots,-1$ & $-2m+1, -2m+3,\dots, 1$\\
 				\hline
 				\textbf{II}& $0$ & $0,1,\dots,m$  & $0,1,\dots,m$ & $3,5,7,\dots, 2m+3$\\
 				\hline
 				\textbf{III}& $1$  & $0,1,\dots,m$  & $m+1,m+2,\dots,2m+1$ & $2m+5, 2m+7, \dots, 4m+5$\\
 				\hline
 				\textbf{IV}& $2,3,\dots$  & $0,1,\dots,m$  & $2m+2,2m+3, \dots,3m+2, 3m+3,\dots, 4m+3, \dots$ & $4m+7, 4m+9, \dots, 6m+7, 6m+9, \dots $\\
 				\hline
 			\end{tabular}
 		\end{center}
 	\end{small}
 \caption{Values of $N$ ($\lambda$, $\mu$) specifying the four regions of the energy spectrum.}
 \label{t1}
 \end{table}

\noindent Inside these four regions, that we report explicitly in Table \ref{t1}, the eigenkets will be then reparametrized in terms of $m$, $\lambda$ and $\mu$. The explicit action of the higher order quantum integrals will then read:
 \begin{align}
 \hat{T}_\pm \ket{N,n_1,n_2} &=t_{\pm}(m; N, n_1, n_2)\ket{N, n_1 \pm (m+1), n_2 \mp (m+1)} \\
 \hat{U}_\pm \ket{N,n_1,n_2} &=u_{\pm}(m; N, n_1, n_2)\ket{N, n_1, n_2 \pm (m+1)}\\
 \hat{V}_\pm \ket{N,n_1,n_2} &=v_{\pm}(m; N, n_1, n_2)\ket{N, n_1 \pm (m+1), n_2} \, ,
 \label{eq:ladder}
 \end{align}
 where $\ket{N,n_1,n_2} \equiv \ket{N(m;\lambda,\mu) ,n_1(m;\lambda,\mu) ,n_2(m;\lambda,\mu)}$ and the functions $(t_\pm , u_\pm, v_\pm)$ will be directly computed from the explicit action of the constituent ladder operators when specific values of $\lambda$ and $\mu$ have been specified for the region we are considering. 
 The action of $\hat{T}_3$ and $\hat{Y}$ on this basis reads:
 \begin{equation}
\hat{T}_3 \ket{N,n_1,n_2}=\frac{n_1-n_2}{2(m+1)} \ket{N, n_1,n_2} \qquad
 \hat{Y} \ket{N,n_1,n_2}=\frac{3n_1+3n_2-2N}{3(m+1)} \ket{N, n_1,n_2} \, .
 \label{eq:cartan}
 \end{equation}
 
\begin{figure}[httb]
 	\begin{center}
 		\begin{tikzpicture}
 		\begin{axis}[ axis lines=center, grid=none,ymin=-2.75, ymax=2.75, xmin=-2.75, xmax=2.75, xlabel={$t_3$},
 	ylabel={$y$}]
 		\addplot[
 		color=cyan,
 		mark=*,
 		mark options={solid},
 		]
 		coordinates {
 			(-13/6,7/9)
 		};
 		\addplot[
 		color=cyan,
 		mark=*,
 		mark options={solid},
 		]
 		coordinates {
 			(-12/6,4/9)
 		};
 		\addplot[
 		color=cyan,
 		mark=*,
 		mark options={solid},
 		]
 		coordinates {
 			(-11/6,1/9)
 		};
 		\addplot[
 		color=cyan,
 		mark=*,
 		mark options={solid},
 		]
 		coordinates {
 			(-10/6,-2/9)
 		};
 		\addplot[
 		color=cyan,
 		mark=*,
 		mark options={solid},
 		]
 		coordinates {
 			(-9/6,-5/9)
 		};
 		\addplot[
 		color=cyan,
 		mark=*,
 		mark options={solid},
 		]
 		coordinates {
 			(-8/6,-8/9)
 		};
 		\addplot[
 		color=cyan,
 		mark=*,
 		mark options={solid},
 		]
 		coordinates {
 			(-7/6,-11/9)
 		};
 		\addplot[
 		color=cyan,
 		mark=*,
 		mark options={solid},
 		]
 		coordinates {
 			(-6/6,-14/9)
 		};
 		\addplot[
 		color=cyan,
 		mark=*,
 		mark options={solid},
 		]
 		coordinates {
 			(-5/6,-17/9)
 		};
 		\addplot[
 		color=cyan,
 		mark=*,
 		mark options={solid},
 		]
 		coordinates {
 			(-4/6,-20/9)
 		};
 		\addplot[
 		color=cyan,
 		mark=*,
 		mark options={solid},
 		]
 		coordinates {
 			(-3/6,-23/9)
 		};
 		\addplot[
 		color=cyan,
 		mark=*,
 		mark options={solid},
 		]
 		coordinates {
 			(7/6,7/9)
 		};
 		\addplot[
 		color=cyan,
 		mark=*,
 		mark options={solid},
 		]
 		coordinates {
 			(-7/6,7/9)
 		};
 		\addplot[
 		color=cyan,
 		mark=*,
 		mark options={solid},
 		]
 		coordinates {
 			(0,-14/9)
 		};
 		\addplot[
 		color=cyan,
 		mark=*,
 		mark options={solid},
 		]
 		coordinates {
 			(5/6,7/9)
 		};
 		\addplot[
 		color=cyan,
 		mark=*,
 		mark options={solid},
 		]
 		coordinates {
 			(6/6,4/9)
 		};
 		\addplot[
 		color=cyan,
 		mark=*,
 		mark options={solid},
 		]
 		coordinates {
 			(-5/6,7/9)
 		};
 		\addplot[
 		color=cyan,
 		mark=*,
 		mark options={solid},
 		]
 		coordinates {
 			(-6/6,4/9)
 		};
 		\addplot[
 		color=cyan,
 		mark=*,
 		mark options={solid},
 		]
 		coordinates {
 			(1/6,-11/9)
 		};
 		\addplot[
 		color=cyan,
 		mark=*,
 		mark options={solid},
 		]
 		coordinates {
 			(-1/6,-11/9)
 		};
 		\addplot[
 		color=cyan,
 		mark=*,
 		mark options={solid},
 		]
 		coordinates {
 			(3/6,7/9)
 		};
 		\addplot[
 		color=cyan,
 		mark=*,
 		mark options={solid},
 		]
 		coordinates {
 			(5/6,1/9)
 		};
 		
 		\addplot[
 		color=cyan,
 		mark=*,
 		mark options={solid},
 		]
 		coordinates {
 			(-3/6,7/9)
 		};
 		\addplot[
 		color=cyan,
 		mark=*,
 		mark options={solid},
 		]
 		coordinates {
 			(-5/6,1/9)
 		};
 		
 		\addplot[
 		color=cyan,
 		mark=*,
 		mark options={solid},
 		]
 		coordinates {
 			(2/6,-8/9)
 		};
 		\addplot[
 		color=cyan,
 		mark=*,
 		mark options={solid},
 		]
 		coordinates {
 			(-2/6,-8/9)
 		};
 		
 		\addplot[
 		color=cyan,
 		mark=*,
 		mark options={solid},
 		]
 		coordinates {
 			(1/6,7/9)
 		};
 		\addplot[
 		color=cyan,
 		mark=*,
 		mark options={solid},
 		]
 		coordinates {
 			(-4/6,-2/9)
 		};
 		
 		\addplot[
 		color=cyan,
 		mark=*,
 		mark options={solid},
 		]
 		coordinates {
 			(4/6,-2/9)
 		};
 		\addplot[
 		color=cyan,
 		mark=*,
 		mark options={solid},
 		]
 		coordinates {
 			(-1/6,7/9)
 		};
 		
 		\addplot[
 		color=cyan,
 		mark=*,
 		mark options={solid},
 		]
 		coordinates {
 			(-3/6,-5/9)
 		};
 		\addplot[
 		color=cyan,
 		mark=*,
 		mark options={solid},
 		]
 		coordinates {
 			(3/6,-5/9)
 		};
 		
 		\addplot[
 		color=cyan,
 		mark=*,
 		mark options={solid},
 		]
 		coordinates {
 			(0,-8/9)
 		};
 		\addplot[
 		color=cyan,
 		mark=*,
 		mark options={solid},
 		]
 		coordinates {
 			(-4/6,4/9)
 		};
 		\addplot[
 		color=cyan,
 		mark=*,
 		mark options={solid},
 		]
 		coordinates {
 			(4/6,4/9)
 		};
 		
 		\addplot[
 		color=cyan,
 		mark=*,
 		mark options={solid},
 		]
 		coordinates {
 			(0,-2/9)
 		};
 		
 		\addplot[
 		color=cyan,
 		mark=*,
 		mark options={solid},
 		]
 		coordinates {
 			(-1/6,1/9)
 		};
 		
 		\addplot[
 		color=cyan,
 		mark=*,
 		mark options={solid},
 		]
 		coordinates {
 			(1/6,1/9)
 		};
 		
 		\addplot[
 		color=cyan,
 		mark=*,
 		mark options={solid},
 		]
 		coordinates {
 			(2/6,-2/9)
 		};
 		\addplot[
 		color=cyan,
 		mark=*,
 		mark options={solid},
 		]
 		coordinates {
 			(-2/6,-2/9)
 		};
 		
 		\addplot[
 		color=cyan,
 		mark=*,
 		mark options={solid},
 		]
 		coordinates {
 			(-1/6,-5/9)
 		};
 		\addplot[
 		color=cyan,
 		mark=*,
 		mark options={solid},
 		]
 		coordinates {
 			(1/6,-5/9)
 		};
 		
 		\addplot[
 		color=cyan,
 		mark=*,
 		mark options={solid},
 		]
 		coordinates {
 			(3/6,1/9)
 		};
 		\addplot[
 		color=cyan,
 		mark=*,
 		mark options={solid},
 		]
 		coordinates {
 			(-3/6,1/9)
 		};
 		
 		\addplot[
 		color=cyan,
 		mark=*,
 		mark options={solid},
 		]
 		coordinates {
 			(2/6,4/9)
 		};
 		\addplot[
 		color=cyan,
 		mark=*,
 		mark options={solid},
 		]
 		coordinates {
 			(-2/6,4/9)
 		};
 		
 		\addplot[
 		color=cyan,
 		mark=*,
 		mark options={solid},
 		]
 		coordinates {
 			(0,4/9)
 		};
 		
 		\end{axis}
 		\end{tikzpicture} \hskip 0.5cm
 		\begin{tikzpicture}
 		\begin{axis}[ axis lines=center, grid=none,ymin=-2.75, ymax=2.75, xmin=-2.75, xmax=2.75, xlabel={$t_3$},
 		ylabel={$y$}]
 		\addplot[
 		color=red,
 		mark=*,
 		mark options={solid},
 		]
 		coordinates {
 			(-13/6,7/9)
 		};
 		\addplot[
 		color=cyan,
 		mark=*,
 		mark options={solid},
 		]
 		coordinates {
 			(-12/6,4/9)
 		};
 		\addplot[
 		color=cyan,
 		mark=*,
 		mark options={solid},
 		]
 		coordinates {
 			(-11/6,1/9)
 		};
 		\addplot[
 		color=red,
 		mark=*,
 		mark options={solid},
 		]
 		coordinates {
 			(-10/6,-2/9)
 		};
 		\addplot[
 		color=cyan,
 		mark=*,
 		mark options={solid},
 		]
 		coordinates {
 			(-9/6,-5/9)
 		};
 		\addplot[
 		color=cyan,
 		mark=*,
 		mark options={solid},
 		]
 		coordinates {
 			(-8/6,-8/9)
 		};
 		\addplot[
 		color=red,
 		mark=*,
 		mark options={solid},
 		]
 		coordinates {
 			(-7/6,-11/9)
 		};
 		\addplot[
 		color=cyan,
 		mark=*,
 		mark options={solid},
 		]
 		coordinates {
 			(-6/6,-14/9)
 		};
 		\addplot[
 		color=cyan,
 		mark=*,
 		mark options={solid},
 		]
 		coordinates {
 			(-5/6,-17/9)
 		};
 		\addplot[
 		color=red,
 		mark=*,
 		mark options={solid},
 		]
 		coordinates {
 			(-4/6,-20/9)
 		};
 		\addplot[
 		color=cyan,
 		mark=*,
 		mark options={solid},
 		]
 		coordinates {
 			(-3/6,-23/9)
 		};
 		\addplot[
 		color=cyan,
 		mark=*,
 		mark options={solid},
 		]
 		coordinates {
 			(7/6,7/9)
 		};
 		\addplot[
 		color=red,
 		mark=*,
 		mark options={solid},
 		]
 		coordinates {
 			(-7/6,7/9)
 		};
 		\addplot[
 		color=cyan,
 		mark=*,
 		mark options={solid},
 		]
 		coordinates {
 			(0,-14/9)
 		};
 		\addplot[
 		color=red,
 		mark=*,
 		mark options={solid},
 		]
 		coordinates {
 			(5/6,7/9)
 		};
 		\addplot[
 		color=cyan,
 		mark=*,
 		mark options={solid},
 		]
 		coordinates {
 			(6/6,4/9)
 		};
 		\addplot[
 		color=cyan,
 		mark=*,
 		mark options={solid},
 		]
 		coordinates {
 			(-5/6,7/9)
 		};
 		\addplot[
 		color=cyan,
 		mark=*,
 		mark options={solid},
 		]
 		coordinates {
 			(-6/6,4/9)
 		};
 		\addplot[
 		color=cyan,
 		mark=*,
 		mark options={solid},
 		]
 		coordinates {
 			(1/6,-11/9)
 		};
 		\addplot[
 		color=red,
 		mark=*,
 		mark options={solid},
 		]
 		coordinates {
 			(-1/6,-11/9)
 		};
 		\addplot[
 		color=cyan,
 		mark=*,
 		mark options={solid},
 		]
 		coordinates {
 			(3/6,7/9)
 		};
 		\addplot[
 		color=cyan,
 		mark=*,
 		mark options={solid},
 		]
 		coordinates {
 			(5/6,1/9)
 		};
 		
 		\addplot[
 		color=cyan,
 		mark=*,
 		mark options={solid},
 		]
 		coordinates {
 			(-3/6,7/9)
 		};
 		\addplot[
 		color=cyan,
 		mark=*,
 		mark options={solid},
 		]
 		coordinates {
 			(-5/6,1/9)
 		};
 		
 		\addplot[
 		color=cyan,
 		mark=*,
 		mark options={solid},
 		]
 		coordinates {
 			(2/6,-8/9)
 		};
 		\addplot[
 		color=cyan,
 		mark=*,
 		mark options={solid},
 		]
 		coordinates {
 			(-2/6,-8/9)
 		};
 		
 		\addplot[
 		color=cyan,
 		mark=*,
 		mark options={solid},
 		]
 		coordinates {
 			(1/6,7/9)
 		};
 		\addplot[
 		color=red,
 		mark=*,
 		mark options={solid},
 		]
 		coordinates {
 			(-4/6,-2/9)
 		};
 		
 		\addplot[
 		color=cyan,
 		mark=*,
 		mark options={solid},
 		]
 		coordinates {
 			(4/6,-2/9)
 		};
 		\addplot[
 		color=red,
 		mark=*,
 		mark options={solid},
 		]
 		coordinates {
 			(-1/6,7/9)
 		};
 		
 		\addplot[
 		color=cyan,
 		mark=*,
 		mark options={solid},
 		]
 		coordinates {
 			(-3/6,-5/9)
 		};
 		\addplot[
 		color=cyan,
 		mark=*,
 		mark options={solid},
 		]
 		coordinates {
 			(3/6,-5/9)
 		};
 		
 		\addplot[
 		color=cyan,
 		mark=*,
 		mark options={solid},
 		]
 		coordinates {
 			(0,-8/9)
 		};
 		\addplot[
 		color=cyan,
 		mark=*,
 		mark options={solid},
 		]
 		coordinates {
 			(-4/6,4/9)
 		};
 		\addplot[
 		color=cyan,
 		mark=*,
 		mark options={solid},
 		]
 		coordinates {
 			(4/6,4/9)
 		};
 		
 		\addplot[
 		color=cyan,
 		mark=*,
 		mark options={solid},
 		]
 		coordinates {
 			(0,-2/9)
 		};
 		
 		\addplot[
 		color=cyan,
 		mark=*,
 		mark options={solid},
 		]
 		coordinates {
 			(-1/6,1/9)
 		};
 		
 		\addplot[
 		color=cyan,
 		mark=*,
 		mark options={solid},
 		]
 		coordinates {
 			(1/6,1/9)
 		};
 		
 		\addplot[
 		color=red,
 		mark=*,
 		mark options={solid},
 		]
 		coordinates {
 			(2/6,-2/9)
 		};
 		\addplot[
 		color=cyan,
 		mark=*,
 		mark options={solid},
 		]
 		coordinates {
 			(-2/6,-2/9)
 		};
 		
 		\addplot[
 		color=cyan,
 		mark=*,
 		mark options={solid},
 		]
 		coordinates {
 			(-1/6,-5/9)
 		};
 		\addplot[
 		color=cyan,
 		mark=*,
 		mark options={solid},
 		]
 		coordinates {
 			(1/6,-5/9)
 		};
 		
 		\addplot[
 		color=cyan,
 		mark=*,
 		mark options={solid},
 		]
 		coordinates {
 			(3/6,1/9)
 		};
 		\addplot[
 		color=cyan,
 		mark=*,
 		mark options={solid},
 		]
 		coordinates {
 			(-3/6,1/9)
 		};
 		
 		\addplot[
 		color=cyan,
 		mark=*,
 		mark options={solid},
 		]
 		coordinates {
 			(2/6,4/9)
 		};
 		\addplot[
 		color=cyan,
 		mark=*,
 		mark options={solid},
 		]
 		coordinates {
 			(-2/6,4/9)
 		};
 		
 		\addplot[
 		color=cyan,
 		mark=*,
 		mark options={solid},
 		]
 		coordinates {
 			(0,4/9)
 		};
 		
 		\begin{scope}[thin,decoration={
 			markings,
 			mark=at position 0.5 with {\arrow{<}}}
 		]

 		\draw[-,  red, postaction={decorate}] (axis cs:	-13/6,7/9) to[bend left=-30] (axis cs:	-7/6,7/9);

 		\draw[-,  red, postaction={decorate}] (axis cs:	-7/6,7/9) to[bend left=-30] (axis cs:	-1/6,7/9);

 		\draw[-,  red, postaction={decorate}] (axis cs:	-1/6,7/9) to[bend left=-30] (axis cs:	5/6,7/9);

 		\draw[-,  red, postaction={decorate}] (axis cs:	5/6,7/9) to[bend left=-30] (axis cs:	2/6,-2/9);

 		\draw[-,  red, postaction={decorate}] (axis cs:-13/6,7/9) to[bend left=-30] (axis cs:-10/6,-2/9);

 		\draw[-,  red, postaction={decorate}] (axis cs:	-10/6,-2/9) to[bend left=-30] (axis cs:-4/6,-2/9);

 		\draw[-,  red, postaction={decorate}] (axis cs:	-4/6,-2/9) to[bend left=-30] (axis cs: 2/6,-2/9);

 		\draw[-,  red,postaction={decorate}] (axis cs:	-10/6,-2/9) to[bend left=-30] (axis cs: -7/6,-11/9);

 		\draw[-,  red, postaction={decorate}] (axis cs:	-7/6,-11/9) to[bend left=-30] (axis cs: -1/6,-11/9);

 		\draw[-,  red, postaction={decorate}] (axis cs:-1/6,-11/9) to[bend left=-30] (axis cs: 2/6,-2/9);

 		\draw[-,  red, postaction={decorate}] (axis cs:	-7/6,-11/9) to[bend left=-30] (axis cs: -4/6,-20/9);

 		\draw[-,  red, postaction={decorate}] (axis cs:	-4/6,-20/9) to[bend left=-30] (axis cs: -1/6,-11/9);

 		\draw[-,  red, postaction={decorate}] (axis cs:	-7/6,7/9) to[bend left=-30] (axis cs: -10/6,-2/9);
 		
 		\draw[-,  red, postaction={decorate}] (axis cs:	-7/6,7/9) to[bend left=-30] (axis cs: -4/6,-2/9);

 		\draw[-,  red, postaction={decorate}] (axis cs:	-4/6,-2/9) to[bend left=-30] (axis cs: -1/6,7/9);

 		\draw[-,  red, postaction={decorate}] (axis cs:	-1/6,7/9) to[bend left=-30] (axis cs: 2/6,-2/9);

 		\draw[-,  red, postaction={decorate}] (axis cs:	-7/6,-11/9) to[bend left=-30] (axis cs:-4/6 ,-2/9);
 		
 		\draw[-,  red, postaction={decorate}] (axis cs:	-4/6,-2/9) to[bend left=-30] (axis cs:-1/6 ,-11/9);
 		
 		\end{scope}
 		
 		\begin{scope}[thin,decoration={
 			markings,
 			mark=at position 0.5 with {\arrow{>}}}
 		]

 		\draw[-, red, postaction={decorate}] (axis cs:	-13/6,7/9) to[bend right=-30] (axis cs:	-7/6,7/9);

 		\draw[-, red, postaction={decorate}] (axis cs:	-7/6,7/9) to[bend right=-30] (axis cs:	-1/6,7/9);

 		\draw[-, red, postaction={decorate}] (axis cs:	-1/6,7/9) to[bend right=-30] (axis cs:	5/6,7/9);

 		\draw[-,  red, postaction={decorate}] (axis cs:	5/6,7/9) to[bend right=-30] (axis cs:	2/6,-2/9);

 		\draw[-,  red, postaction={decorate}] (axis cs:-13/6,7/9) to[bend right=-30] (axis cs:-10/6,-2/9);

 		\draw[-,  red, postaction={decorate}] (axis cs:	-10/6,-2/9) to[bend right=-30] (axis cs:-4/6,-2/9);

 		\draw[-,  red, postaction={decorate}] (axis cs:	-4/6,-2/9) to[bend right=-30] (axis cs: 2/6,-2/9);

 		\draw[-,  red, postaction={decorate}] (axis cs:	-10/6,-2/9) to[bend right=-30] (axis cs: -7/6,-11/9);

 		\draw[-,  red, postaction={decorate}] (axis cs:	-7/6,-11/9) to[bend right=-30] (axis cs: -1/6,-11/9);

 		\draw[-,  red, postaction={decorate}] (axis cs:	-1/6,-11/9) to[bend right=-30] (axis cs: 2/6,-2/9);

 		\draw[-,  red, postaction={decorate}] (axis cs:	-7/6,-11/9) to[bend right=-30] (axis cs: -4/6,-20/9);

 		\draw[-,  red, postaction={decorate}] (axis cs:	-4/6,-20/9) to[bend right=-30] (axis cs: -1/6,-11/9);
 		
 		\draw[-,  red, postaction={decorate}] (axis cs:	-7/6,7/9) to[bend right=-30] (axis cs: -10/6,-2/9);

 		\draw[-,  red, postaction={decorate}] (axis cs:	-7/6,7/9) to[bend right=-30] (axis cs: -4/6,-2/9);
 		
 		\draw[-,  red, postaction={decorate}] (axis cs:	-4/6,-2/9) to[bend right=-30] (axis cs: -1/6,7/9);

 		\draw[-,  red, postaction={decorate}] (axis cs:	-1/6,7/9) to[bend right=-30] (axis cs: 2/6,-2/9);

 		\draw[-,  red, postaction={decorate}] (axis cs:	-7/6,-11/9) to[bend right=-30] (axis cs:-4/6 ,-2/9);

 		\draw[-,  red,postaction={decorate}] (axis cs:	-4/6,-2/9) to[bend right=-30] (axis cs:-1/6 ,-11/9);
 		
 		\end{scope}

 		\begin{scope}[thin,decoration={
 			markings,
 			mark=at position 0.5 with {\arrow{>}}}
 		] 
 		\draw[-, red, postaction={decorate}] (axis cs:	1,2) to[bend right=-30] (axis cs:2,2);
 		\draw[-, red, postaction={decorate}] (axis cs:	2,2) to[bend right=-30] (axis cs:1.5,1);
 		\draw[-, red, postaction={decorate}] (axis cs:	1.5,1) to[bend right=-30] (axis cs:1,2);
 		\end{scope}
 		
 		\begin{scope}[thin,decoration={
 			markings,
 			mark=at position 0.5 with {\arrow{<}}}
 		] 
 		\draw[-, red, postaction={decorate}] (axis cs:	1,2) to[bend left=-30] (axis cs:2,2);
 		\draw[- ,red, postaction={decorate}] (axis cs:	2,2) to[bend left=-30] (axis cs:1.5,1);
 		\draw[-, red, postaction={decorate}] (axis cs:	1.5,1) to[bend left=-30] (axis cs:1,2);
 		\end{scope}
 		
 		\newcommand{\plus}{\raisebox{.4\height}{\scalebox{.4}{$\pm$}}}
 		\newcommand{\minus}{\raisebox{.4\height}{\scalebox{.4}{$\mp$}}}
 		\begin{scope}
 		\node[] at (axis cs: 1.5,1.975) {$\text{\tiny{T}}_{\text{\plus}}$};
 		\node[] at (axis cs: 1.75,1.45) {$\text{\tiny{V}}_{\text{\plus}}$};
 		\node[] at (axis cs: 1.275,1.45) {$\text{\tiny{U}}_{\text{\minus}}$};
 		\end{scope}
 		\end{axis}
 		\end{tikzpicture}
 	\end{center}
 	\caption{On the left, a typical triangular lattice representing physical states in the plane $(t_3,y)$ is given, for a fixed value of $N(m,\lambda^*,\mu^*)=N(2;0,2)=7$. On the right, we show an example of the action of the higher order quantum integrals on a subset of connected states (multiplet).}
 	\label{fig1}
 \end{figure}
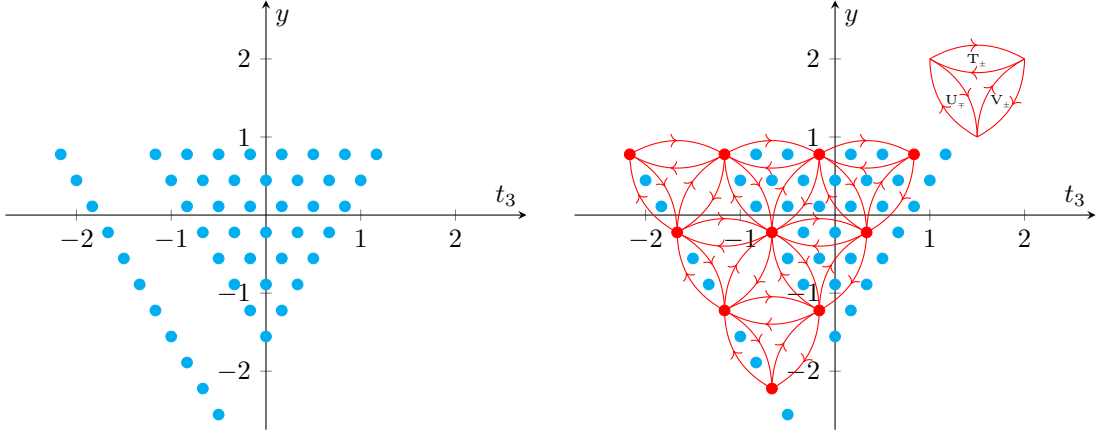

 \noindent To have an idea on how higher order quantum integrals connect states, at fixed energy level $N(m,\lambda^*,\mu^*)=(m+1)\lambda^*+\mu^*$, it is convenient to rely on a graphical representation of their action in the plane $(t_3, y)$. In this plane physical states, as it can be understood from \eqref{eq:cartan},  define triangular lattices. An illustrative example is reported in \figref{fig1}.
 
\noindent Throughout the paper, for the sake of clarity, we will be representing with the same color physical states that are connected by the action of the higher order quantum integrals. Then, the construction proceeds as follows:
 \begin{enumerate}
\item Inside a specific region of the energy spectrum, suppose a number $K=K(m; N)$ of basis eigenstates of $\hat{H}$ arise.  We reparametrise these physical states in the basis $\ket{N, n_1, n_2}$ in terms of appropriate values of $\lambda$ and $\mu$ valid inside the region we are considering. This specify the following set composed by $K=K(m; N(m;\lambda,\mu))$ eigenkets:
 \begin{equation}
\small \mathcal{S}(m;\lambda, \mu):=\bigl\{\ket{N(m;\lambda,\mu), n_1(m;\lambda,\mu), n_2(m;\lambda,\mu)}^{[1]}, \dots, \ket{N(m;\lambda,\mu), n_1(m;\lambda,\mu), n_2(m;\lambda,\mu)}^{[\#\mathcal{S}(m;\lambda, \mu)]}\bigl\} \, ,
\label{eq:kets}
\end{equation}
 where we indicated $K \equiv \#\mathcal{S}(m;\lambda, \mu)$, the cardinality of the set \eqref{eq:kets}.
 \item At fixed $\lambda=\lambda^*$, we decompose the above set into $m+1$ subsets specified by values of $\mu=0,1,\dots, m$, i.e.:
  \begin{equation}
 \mathcal{S}(m;\lambda^*, \mu):=\mathcal{S}(m;\lambda^*, 0) \cup \mathcal{S}(m;\lambda^*, 1) \cup \dots \cup \mathcal{S}(m;\lambda^*, m) \, .
 \label{eq:subkets}
 \end{equation}
 Each subset $\mathcal{S}(m;\lambda^*, \mu^*)$, for fixed $\mu=\mu^*$, will then contain $\#\mathcal{S}(m;\lambda^*, \mu^*)$ states and, of course:  $$\sum_{j=0}^{m}\#\mathcal{S}(m;\lambda^*, j)=\#\mathcal{S}(m;\lambda^*, \mu) \, .$$
\item We impose the conditions: 
 \begin{equation}
 \begin{cases}
\hat{T}_+ \ket{N(m;\lambda^*,\mu), n_1(m;\lambda^*,\mu), n_2(m;\lambda^*,\mu)}^{[j]} = 0\\  \hat{V}_+ \ket{N(m;\lambda^*,\mu), n_1(m;\lambda^*,\mu), n_2(m;\lambda^*,\mu)}^{[j]} =0 
\end{cases} \qquad \text{for each} \quad j = 1, \dots, \# \mathcal{S}(m;\lambda^*, \mu)=0
\label{eq:conditionshw}
\end{equation}
where $\ket{N(m;\lambda^*,\mu), n_1(m;\lambda^*,\mu), n_2(m;\lambda^*,\mu)}^{[j]} \in \mathcal{S}(m;\lambda^*, \mu)$, namely we look at those states that are simultaneously annihilated by the action of two raising operators. These actions, that can be constructed from the action of the constituent ladder operators in the three axes as specified previously, when restricted to the specific regions ($\lambda^*$, $\mu=0,1,\dots, m$), allow us to find the values $n_1=n_1(\lambda^*,\mu)$ and $n_2=n_2(\lambda^*,\mu)$ for which the system of equations is satisfied inside each specific subset specified by $\mu=\mu^*$. This operation leads us to identify, for each fixed $N(m;\lambda^*, \mu^*)=(m+1)\lambda^*+\mu^*$, a number $\#\tilde{\mathcal{S}}(m;\lambda^*, \mu^*)$ of annihilated states:
\begin{equation}
\footnotesize \tilde{\mathcal{S}}(m;\lambda^*, \mu^*)=\bigl\{\ket{N(m;\lambda^*,\mu^*), n_1(m;\lambda^*,\mu^*), n_2(m;\lambda^*,\mu^*)}^{[1]}, \dots, \ket{N(m;\lambda^*,\mu^*), n_1(m;\lambda^*,\mu^*), n_2(m;\lambda^*,\mu^*)}^{[\#\tilde{\mathcal{S}}(m;\lambda^*, \mu^*)]}\bigl\} \, .
\label{eq:highestketsfixeden}
\end{equation}
 In the following, to simplify the notation, we indicate \eqref{eq:conditionshw} as an operation on sets, i.e.: $$(\hat{T}_+,\hat{V}_+) \rhd \mathcal{S}(m;\lambda^*, \mu)=0 \, .$$
 \item We impose the condition: 
 \begin{equation}
 (\hat{V}_-)^{p+1} \ket{N(m;\lambda^*,\mu^*), n_1(m;\lambda^*,\mu^*), n_2(m;\lambda^*,\mu^*)}^{[j]} =0 \quad \text{for} \quad j =1, \dots, \#\tilde{\mathcal{S}}(m;\lambda^*, \mu^*) 
 \label{eq:cond}
 \end{equation}
 where $\ket{N(m;\lambda^*,\mu^*), n_1(m;\lambda^*,\mu^*), n_2(m;\lambda^*,\mu^*)}^{[j]} \in \tilde{\mathcal{S}}(m;\lambda^*, \mu^*)$. This allow us to determine the values $p_k$, $k=1,\dots, \#\tilde{\mathcal{S}}(m;\lambda^*, \mu^*)$, i.e. to determine the boundary of the triangular multiplets composed by connected states. We indicate the sets composed by these connected states as $\mathcal{P}_k(m;\lambda^*, \mu^*)$. Each set $\mathcal{P}_{k^*}(m;\lambda^*, \mu^*)$ turns out to have cardinality  $\#\mathcal{P}_{k^*}(m;\lambda^*, \mu^*)=(p_{k^*}+1)(p_{k^*}+2)/2$.  
 In the following, we will specify these multiplets in term of their dimensions $\boldsymbol{p_{k^*}}=(p_{k^*}+1)(p_{k^*}+2)/2$, so that depending on the value assumed by $p_{k^*}$ we can have, besides singlets, $\mathbf{3}$, $\mathbf{6}$, $\mathbf{10}$, $\dots$ multiplets of states connected by the action of the higher order quantum integrals of motion. 
This means that, at fixed energy, i.e. fixed $N=(m+1)\lambda^*+\mu^*$ the set of eigenstates:
\begin{equation}
\footnotesize \mathcal{S}(m;\lambda^*, \mu^*)=\bigl\{\ket{N(m;\lambda^*,\mu^*), n_1(m;\lambda^*,\mu^*), n_2(m;\lambda^*,j^*)}^{[1]}, \dots, \ket{N(m;\lambda^*,\mu^*), n_1(m;\lambda^*,\mu^*), n_2(m;\lambda^*,\mu^*)}^{[\#\mathcal{S}(m;\lambda^*, \mu^*)]}\bigl\}
\label{eq:eigenset}
\end{equation}
will be organized into $\#\tilde{\mathcal{S}}(m;\lambda^*, \mu^*)$ multiplets of dimension $(p_{k}+1)(p_{k}+2)/2$, for $k=1, \dots, \#\tilde{\mathcal{S}}(m;\lambda^*, \mu^*)$, namely:
$$\mathcal{S}(m;\lambda^*, \mu^*) = \bigcup_{k=1}^{\#\tilde{\mathcal{S}}(m;\lambda^*, \mu^*)} \mathcal{P}_{k}(m;\lambda^*, \mu^*)$$
with:
 $$\#\mathcal{S}(m;\lambda^*, \mu^*)=\sum_{k=1}^{\#\tilde{\mathcal{S}}(m;\lambda^*, \mu^*)}(p_{k}+1)(p_{k}+2)/2 \, .$$

 \item Once multiplets have been identified, one can construct their \textquotedblleft spectrum\textquotedblright  by applying:
 \begin{equation}
 (\hat{T}_-)^{a_2}(\hat{V}_-)^{a_1} \ket{N(m;\lambda^*,\mu^*), n_1(m;\lambda^*,\mu^*), n_2(m;\lambda^*,\mu^*)}^{[j]} \quad \text{for}\quad  j=1, \dots, \#\mathcal{P}_{k^*}(m;\lambda^*, \mu^*)
 \label{eq:multiplet}
 \end{equation}
 with $0 \leq a_1 \leq p_{k^*}$ and $0 \leq a_2 \leq p_{k^*}-a_1$, where $\ket{N(m;\lambda^*,\mu^*), n_1(m;\lambda^*,\mu^*), n_2(m;\lambda^*,\mu^*)}^{[j]} \in \mathcal{P}_{k^*}(m;\lambda^*, \mu^*)$. 
 
At a fixed $N=(m+1)\lambda^*+\mu^*$ each state $\ket{N(m;\lambda^*,\mu^*), n_1(m;\lambda^*,\mu^*), n_2(m;\lambda^*,\mu^*)}^{[j^*]} \in \mathcal{P}_{k^*}(m;\lambda^*, \mu^*)$  belonging to a given multiplet will be characterized by the values ($N,\boldsymbol{p_{k^*}}$) which provide information on the energy and on the type of multiplet it belongs to, respectively. Then, the pair of values $(t_3, y)$ is used to specify their location inside a given multiplet, in particular their rescaled values $(\tilde{t}_3, \tilde{y})$ are considered to set the center of gravity of the various triangles characterizing each irrep in the origin of the $(t_3,y)$ plane. Finally, when other multiplets are considered for the same values of $N$, another label $\tau$ is necessary to discriminate among repeated irreps arising with the same $\boldsymbol{p_{k^*}}$. Clearly, normalizations of these states have to be taken into account in the construction.
 
 \item The procedure terminates when for each specific region such that $N=(m+1)\lambda+\mu$ all $\#\mathcal{S}(m;\lambda, \mu)$ states in \eqref{eq:kets} are organized into multiplets of connected states, so that irreps ($N, \mathbf{p}$) and their basis states have been completely specified. 
 \end{enumerate}

\noindent From now on, our analysis will be performed inside each of the four specific regions we introduced. In particular, our main aim is restricted to the characterization, for different values of $N$ (i.e. $\lambda$ and $\mu$), of the explicit patterns of the $\mathbf{p}$ multiplets. This allows us to provide the total number of irreps per given energy level and to characterize the degeneracies of this model. We should again remark, however, that this is not the algebra spanned by the minimal order generators for this model. This means that states that are not connected by the action of the higher order operators might be actually connected by the action of the existing first-order ones. Therefore, even if this represents a good starting point to deal with these kind of models in three dimensions, the polynomial algebra we introduced arises more naturally when at least two rationally extended oscillators (in two different axes) are considered, so that no residual polar symmetry (leading to first-order quantum integrals) is preserved.

 \subsubsection*{Region I}
\begin{figure}[h!]
	\begin{center}
	\begin{tikzpicture}
		\begin{axis}[ axis lines=center, grid=none,ymin=-1.1, ymax=1.1, xmin=-1.1, xmax=1.1, xlabel={$t_3$},
		ylabel={$y$}]
\addplot[
color=cyan,
mark=*,
mark options={solid},
]
coordinates {
	(-3/6,-3/9)
};
		\end{axis}
		\end{tikzpicture} \hskip 0.5 cm
			\begin{tikzpicture}
		\begin{axis}[ axis lines=center, grid=none,ymin=-1.1, ymax=1.1, xmin=-1.1, xmax=1.1, xlabel={$t_3$},
		ylabel={$y$}]
		\addplot[
		color=cyan,
		mark=*d,
		mark options={solid},
		]
		coordinates {
			(-3/6,-5/9)
		};
		\addplot[
	color=cyan,
	mark=*,
	mark options={solid},
	]
	coordinates {
		(-4/6,-2/9)
	};
		\addplot[
color=cyan,
mark=*,
mark options={solid},
]
coordinates {
	(-3/6,-5/9)
};
		\end{axis}
		\end{tikzpicture} \hskip 0.5 cm
		\begin{tikzpicture}
		\begin{axis}[ axis lines=center, grid=none,ymin=-1.1, ymax=1.1, xmin=-1.1, xmax=1.1, xlabel={$t_3$},
		ylabel={$y$}]
		\addplot[
		color=cyan,
		mark=*,
		mark options={solid},
		]
		coordinates {
			(-3/6,-7/9)
		};
		\addplot[
	color=cyan,
	mark=*,
	mark options={solid},
	]
	coordinates {
		(-4/6,-4/9)
	};
	\addplot[
color=cyan,
mark=*,
mark options={solid},
]
coordinates {
	(-5/6,-1/9)
};
		\end{axis}
		\end{tikzpicture}
			\caption{Physical states arising for $N=-3$ (upper left), $N=-2$ (upper right) and $N=-1$ ($m=2$).}
				\label{fig2}
				\end{center}
\end{figure}
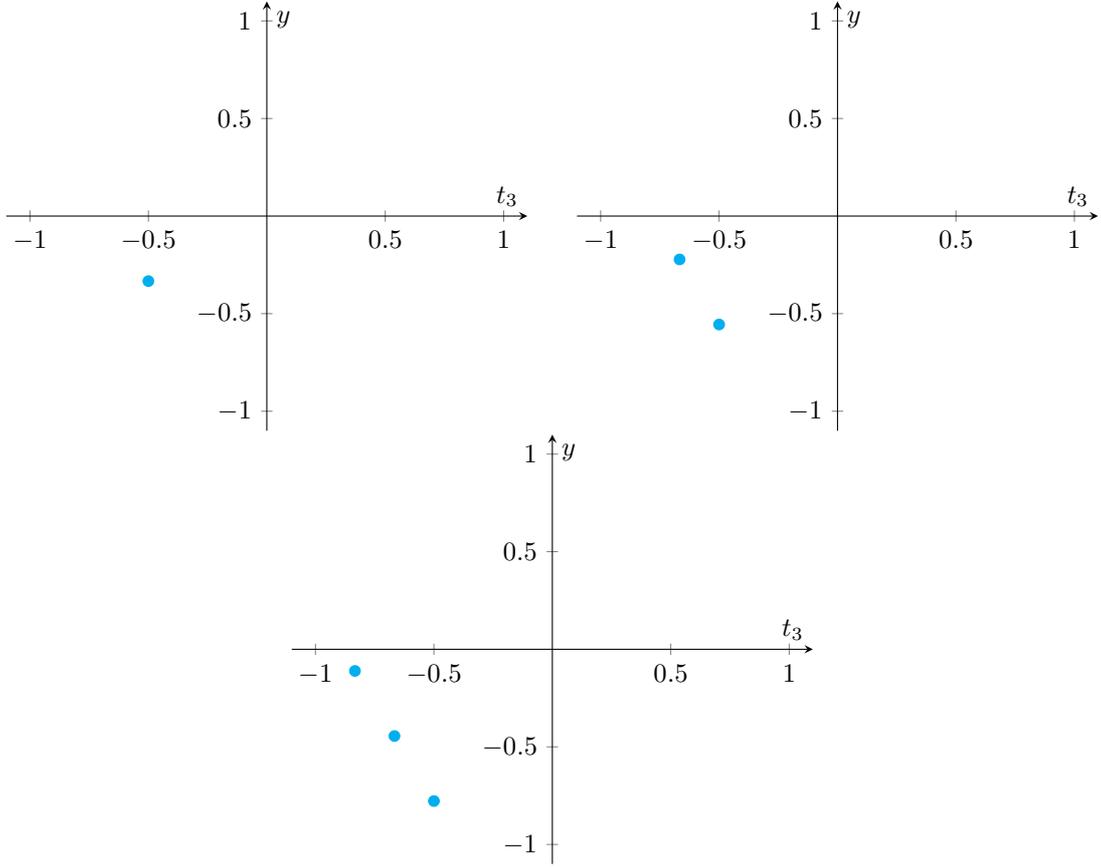

\noindent In region I, i.e. for $\lambda=-1$ and $\mu=0,1,\dots, m$, the basis states can be parametrized as:
\begin{equation*}\ket{N, n_1, n_2} \qquad \text{with} \quad N=\mu-m-1,\quad  n_1=-m-1 \quad \text{and} \quad 0 \leq n_2 \leq \mu \, .
\end{equation*}
We thus have the following set of states:
$$\mathcal{S}(m; -1, \mu)=\{\ket{\mu-m-1, -m-1,n_2}\} \, , \quad   (0 \leq n_2 \leq \mu) \, .$$
 If we decompose this set as $\mathcal{S}(m; -1, \mu) = \mathcal{S}(m; -1, 0) \cup \mathcal{S}(m; -1, 1) \cup \dots \cup \mathcal{S}(m; -1, m)$, namely: 
\begin{equation}
\small \mathcal{S}(m; -1, \mu) = \{\ket{-m-1,-m-1,0}\} \cup \{\ket{-m,-m-1,0},\ket{-m,-m-1,1}\}\cup \dots \cup \{\ket{-1,-m-1,0}, \dots, \ket{-1,-m-1,m}\}
\label{eq:subdec}
\end{equation}

\noindent its cardinality is easily computed to be $\# \mathcal{S}(m; -1, \mu)=1+2+\dots +m+1=\sum_{j=1}^{m+1} j = (m+1)(m+2)/2$. This represents the total number of basis states arising in this region. All these states are annihilated by the action of the higher order integrals.  For $N=-m-1$, i.e. $\mu=0$, there is just one state (a singlet) which we indicate as $\mathbf{1}$. For $N=-m$, i.e. $\mu=1$, there are two singlets, so that we can indicate them as $\mathbf{1}^2$. The process continues until we reach $N=-1$, i.e. $\mu = m$, where we have $\mathbf{1}^{m+1}$. 
\noindent For each even $m$, the eigenvalues of $(\hat{T}_3, \hat{Y})$ operators associated to these states turn out to be:
\begin{equation}
t_3(m; n_2) = -\frac{m+1+n_2}{2(m+1)} \, , \quad y(m; \mu, n_2) = \frac{3 n_2- (2\mu +m+1)}{3(m+1)} \qquad (0\leq n_2 \leq \mu) \, .
\label{eq:t3y}
\end{equation}
These can be plotted in the plane $(t_3, y)$ for different values of $\mu=0,1,\dots, m$ at fixed $m$ (see \figref{fig2}).

\noindent For different values of $m$, the obtained results for region I are summarised in Table \ref{t2}.
\begin{table}[h!]
	\begin{small}
		\begin{center}
			\begin{tabular}{|c c c c c|} 
				\hline
				$\lambda$ & $\mu$ & $\mathbf{p}$ multiplets & Number of Irreps per level & \text{deg}($E_N$)  \\ [0.1ex] 
				\hline
				$-1$ & $0,1,2,\dots,m$ & $\boldsymbol{1}^{\mu+1}$ & $\mu+1$ & $\mu+1$
				\\ [1ex] 
				\hline
			\end{tabular}
		\end{center}
	\end{small}
\caption{$\mathbf{p}$  multiplets with their number of occurrences, number of irreps of the polynomial algebra and degeneracy per energy level characterizing region I.}
\label{t2}
\end{table}

\newpage \subsubsection*{Region II}
\noindent With the reparametrisation $N=(m+1)\lambda+\mu$, we see that states obtained for $\lambda =0$ and $\mu=0,1 \dots, m$ can be parametrized as: 
\begin{equation}
\ket{N, n_1, n_2} \qquad \text{with} \quad N=\mu,\quad  n_1=-m-1, 0, \dots, \mu \quad \text{and} \quad 0 \leq n_2 \leq \mu-n_1 \quad  (\mu=0,1,\dots, m) \, .
\label{statesII}
\end{equation}

\noindent Explictly, we have the following set of states:

\begin{small}
\begin{center}
$\mathcal{S}(m; 0, \mu) = \begin{Bmatrix}
\ket{\mu, -m-1,0} & \ket{\mu, 0, 0} & \ket{\mu, 1, 0} & \dots & \ket{\mu,\mu-2,0}  & \ket{\mu,\mu-1,0} & \ket{\mu,\mu,0}\\
\ket{\mu, -m-1,1} & \ket{\mu, 0,1} & \ket{\mu,1,1} & \dots & \ket{\mu,\mu-2,1} &  \ket{\mu,\mu-1,1} & \\ 
\ket{\mu, -m-1,2} & \ket{\mu, 0,2} & \ket{\mu,1,2} & \dots & \ket{\mu, \mu-2,2} &   & \\ 
\vdots & \vdots & \vdots &  &  &   &  \\
\ket{\mu, -m-1,\mu+m-1} & \ket{\mu, 0,\mu+m-1} & \ket{\mu,1,\mu+m-1} &  &  &   & \\ 
\ket{\mu, -m-1,\mu+m} & \ket{\mu, 0,\mu+m} &  &  &  &  & \\ 
\ket{\mu, -m-1,\mu+m+1} &  &  &  &  &   & 
\end{Bmatrix}$
\end{center}
\end{small}

\noindent At fixed $\mu=\mu^*$, the are a total number of $(\mu^*+2)(\mu^*+3)/2+m$ states. Thus, in this region, the total number of basis states is $\sum_{j=0}^m\bigl((j+2)(j+3)/2+m\bigl)=(m+1)(m^2+14m+18)/6$. 
 \noindent  Starting from the set of states $\mathcal{S}(m;0,\mu)$, we first decompose it as:
 
  $$\mathcal{S}(m;0,\mu)=\mathcal{S}(m;0,0) \cup \mathcal{S}(m;0,1) \cup \dots \cup  \mathcal{S}(m;0,m) \, ,$$ i.e.:

	\begin{small}
		\begin{center}
			$\mathcal{S}(m;0, \mu) = \begin{Bmatrix}
				\ket{0, -m-1,0} & \ket{0, 0, 0} \\
				\ket{0, -m-1,1} &   \\ 
				\ket{0, -m-1,2} &   \\ 
				\vdots &    \\
				\ket{0, -m-1,m-1}&  \\ 
				\ket{0, -m-1,m} &   \\ 
				\ket{0, -m-1,m+1} & \\ 
		\end{Bmatrix}$
		$\cup$ $\begin{Bmatrix} 
			\ket{1, -m-1,0} & \ket{1, 0, 0} & \ket{1,1,0} \\
			\ket{1, -m-1,1} & \ket{1, 0, 1}  &\\ 
			\ket{1, -m-1,2} &  & \\ 
			\vdots &  &  \\
			\ket{1, -m-1,m} & & \\ 
			\ket{1, -m-1,m+1} &   &\\ 
			\ket{1, -m-1,m+2} & &\\ 
		\end{Bmatrix}
		\cup \dots \cup
	\begin{Bmatrix}
	\ket{m, -m-1,0} & \ket{m, 0, 0} & \ket{m, 1, 0} & \dots & \ket{m,m-2,0}  & \ket{m,m-1,0} & \ket{m,m,0}\\
	\dots & \ket{m,m-2,1} &  \ket{m,m-1,1} & \\ 
	\ket{m, -m-1,2} & \ket{m, 0,2} & \ket{m,1,2} & \dots & \ket{m, m-2,2} &   & \\ 
	\vdots & \vdots & \vdots &  &  &   &  \\
	\ket{m, -m-1,2m-1} & \ket{m, 0,2m-1} & \ket{m,1,2m-1} &  &  &   & \\ 
	\ket{m, -m-1,2m} & \ket{m, 0,2m} &  &  &  &  & \\ 
	\ket{m, -m-1,2m+1} &  &  &  &  &   & \\ [1ex] 
	\end{Bmatrix}$
		\end{center}
	\end{small}
 \noindent Then, we impose: $$(\hat{T}_+,\hat{V}_+) \rhd \{\mathcal{S}(m;0,0),\mathcal{S}(m;0,1), \dots, \mathcal{S}(m;0,m)\}=0 \, .$$  In this way, we select the following subset composed by $\# \tilde{\mathcal{S}}(m;0,\mu)=(m+1)(m^2+8m+6)/6$ states:
\begin{small}
	\begin{center}
			$\tilde{\mathcal{S}}(m;0, \mu) = \begin{Bmatrix}
				 & \ket{0, 0, 0} \\
				\ket{0, -m-1,1} &   \\ 
				\ket{0, -m-1,2} &   \\ 
				\vdots &    \\
				\ket{0, -m-1,m-1} &  \\ 
				\ket{0, -m-1,m} &   
			\end{Bmatrix}$
			$\cup$ $\begin{Bmatrix}
				 & \ket{1, 0, 0} & \ket{1,1,0}  \\
				 &   \ket{1,0,1} &\\ 
				\ket{1, -m-1,2} &  & \\ 
				\vdots &  &  \\
				\ket{1, -m-1,m} & & 
			\end{Bmatrix}$
			$\cup \dots \cup$ 
			$\begin{Bmatrix}
				 & \ket{m, 0, 0} & \ket{m, 1, 0} & \dots &  \ket{m,m-2,0} & \ket{m,m-1,0} & \ket{m,m,0}\\
				 & \ket{m, 0,1} & \ket{m,1,1} & \dots & \ket{m,m-2,1} &  \ket{m,m-1,1} & \\ 
				 & \ket{m, 0,2} & \ket{m,1,2} & \dots & \ket{m, m-2,2} &   & \\ 
				\vdots & \vdots & \vdots & \iddots &  &   &  \\
				& \ket{m, 0,m-1} & \ket{m,1,m-1} &  &  &   & \\ 
				& \ket{m, 0, m} &  &  &  &  & 
			\end{Bmatrix}$ \, .
		\end{center}
	\end{small}

\noindent At this point, by imposing: $$(\hat{V}_-)^{p+1} \rhd \{\tilde{\mathcal{S}}(m;0,0),\tilde{\mathcal{S}}(m;0,1), \dots, \tilde{\mathcal{S}}(m;0,m)\}=0\, ,$$ to each of these subsets, we find the values of $p$ associated to the various states, so that we are finally able to organize them into $\mathbf{p}$ multiplets. For the sake of clarity, let us focus on some specific examples. 
\noindent For $\mu=0$, the following $m+3$ states arise:
\begin{equation}
\mathcal{S}(m;0,0)=\{\ket{0, -m-1, 0} \, \ket{0, -m-1, 1}  \, , \ket{0, -m-1, 2}, \dots, \ket{0, -m-1, m+1} , \ket{0,0,0}\}
\label{eq:phys0} \, .
\end{equation}

\noindent Among these states, we select the ones annihilated by $(\hat{T}_+, \hat{V}_+)$. The action of $\hat{T}_+$ specifies the following values of $n_2$, $n_2=0,1, \dots, m$. Thus, the set of $m+3$ states reduces to the following set composed by $m+2$ states:
\begin{equation}
\{\ket{0, -m-1, 0} \, \ket{0, -m-1, 1}  \, , \ket{0, -m-1, 2}, \dots,\ket{0, -m-1, m},  \ket{0,0,0}\}
\label{eq:phys0tp} \, .
\end{equation}
Among these states, the following $m+1$ are also annihilated by the action of $\hat{V}_+$:
\begin{equation}
\tilde{\mathcal{S}}(m;0,0) = \{\ket{0, -m-1, 1}  \, , \ket{0, -m-1, 2}, \dots,\ket{0, -m-1, m},  \ket{0,0,0}\} \, .
\label{anni}
\end{equation}
This is because this operator annihilates states for which $n_1+n_2 = 0,-1,-2, \dots, -m$ in this region for $\mu=0$. 
Therefore, the states annihilated by the action of both operators turn out to be:
\begin{equation}
\ket{0, n_1, n_2}\quad  \text{with} \quad n_1= -m-1 \, ,  n_2=1, \dots, m \quad \text{or}\quad  n_1= n_2=0 \, . 
\label{eq:phys0t} 
\end{equation}

\begin{rem}
In general, for $N=\mu$, starting from the set of basis states \eqref{statesII}, the ones annihilated by $(\hat{T}_+, \hat{V}_+)$ are those for which the following systems of equations is satisfied:
\begin{equation}
\begin{cases}
n_1+ n_2 =\mu, \mu-1, \dots, \mu-m \\
n_2 =0, 1,\dots, m \, .
\end{cases}
\label{eq:cases1}
\end{equation}
This means that at fixed $\mu=\mu^*$, for different values of $n_1=n_1^*, n_1^{**}, \dots$ we obtain the set of values of $n_2$ from both equations. The intersection set of these values give us the solution we are looking for. The are a total number of $\# \tilde{\mathcal{S}}(m;0,\mu)=(m+1)(5m^2+m+6)/6 $ of these states.
\end{rem}
\noindent At this point, by applying the operator $\hat{V}_-$ a repeated number of times until we get zero, we can specify  the multiplets composed by connected states. In particular, for the above set we get:
\begin{equation}
\begin{cases}
\hat{V}_- \ket{0, -m-1, j} = 0 \hskip 1cm \text{for} \quad j=1,\dots, m\\
(\hat{V}_-)^2 \ket{0,0,0} = 0 \, .
\end{cases}
\label{eq:multi}
\end{equation}
%
\noindent This means that the $m+3$ states organize as one triplet $\mathbf{3}$ and $m$ singlets $\mathbf{1}^m$ under the action of the higher order integrals. If $m=2$, for example, five physical states arise:
$$\mathcal{S}(2;0,0)=\{\ket{0,-3,0}, \ket{0,-3,1},\ket{0,-3,2},\ket{0,-3,3},\ket{0,0,0}\} \, .$$

\noindent The states annihilated by the action of $(\hat{T}_+,\hat{V}_+)$ are $\tilde{S}(2, 0,0)=\{\ket{0,-3,1}, \ket{0,-3,2}, \ket{0,0,0}\}$. The operator $\hat{V}_-$ annihilates the first two, so they can be considered as singlets, whereas for the third one we have $(\hat{V}_-)^2\ket{0,0,0} =0$. At this point, by computing the explicit action of ladder operators on $\ket{N, n_1, n_2}=\ket{0,0,0}$ we get the path:
\begin{equation} 
\hat{T}_+\hat{U}_+\hat{V}_-\ket{0,0,0} \propto \ket{0,0,0} \, .
\label{eq:ope}
\end{equation}

\begin{figure}[httb]
	\begin{center}
		\begin{tikzpicture}
		\begin{axis}[ axis lines=center, grid=none,ymin=-1.1, ymax=1.1, xmin=-1.1, xmax=1.1, xlabel={$t_3$},
		ylabel={$y$}]
		\addplot[
		color=cyan,
		mark=*,
		mark options={solid},
		]
		coordinates {
			(-1,0)
		};
		\addplot[
		color=cyan,
		mark=*,
		mark options={solid},
		]
		coordinates {
			(-5/6,-1/3)
		};
		\addplot[
		color=cyan,
		mark=*,
		mark options={solid},
		]
		coordinates {
			(-2/3,-2/3)
		};
		\addplot[
		color=cyan,
		mark=*,
		mark options={solid},
		]
		coordinates {
			(-1/2,-1)
		};
		\addplot[
		color=cyan,
		mark=*,
		mark options={solid},
		]
		coordinates {
			(0,0)
		};
		\end{axis}
		\end{tikzpicture} \hskip 0.5cm
			\begin{tikzpicture}
		\begin{axis}[ axis lines=center, grid=none,ymin=-1.1, ymax=1.1, xmin=-1.1, xmax=1.1, xlabel={$t_3$},
		ylabel={$y$}]
		\addplot[
		color=red,
		mark=*,
		mark options={solid},
		]
		coordinates {
			(-1,0)
		};
		\addplot[
		color=black,
		mark=x,
		mark options={solid},
		]
		coordinates {
			(-5/6,-1/3)
		};
		\addplot[
		color=black,
		mark=x,
		mark options={solid},
		]
		coordinates {
			(-2/3,-2/3)
		};
		\addplot[
		color=red,
		mark=*,
		mark options={solid},
		]
		coordinates {
			(-1/2,-1)
		};
		\addplot[
		color=red,
		mark=*,
		mark options={solid},
		]
		coordinates {
			(0,0)
		};
		\end{axis}
		\end{tikzpicture}
	\end{center}
	\caption{Physical states arising for $N=0$ represented in the plane $(t_3,y)$ for $m=2$. States connected by the action of higher order integrals are indicated in red.}
	\label{fig3}
\end{figure}
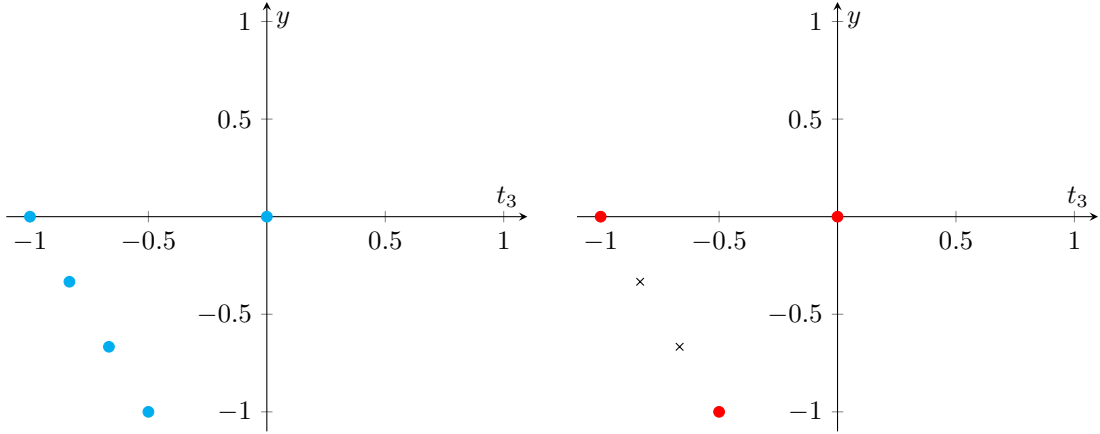

\noindent We represent this situation in \figref{fig3}, where we indicate in red states belonging to the same $\mathbf{3}$, i.e. states connected by the action of the higher order quantum integrals. The remaining states are the two singlets. 

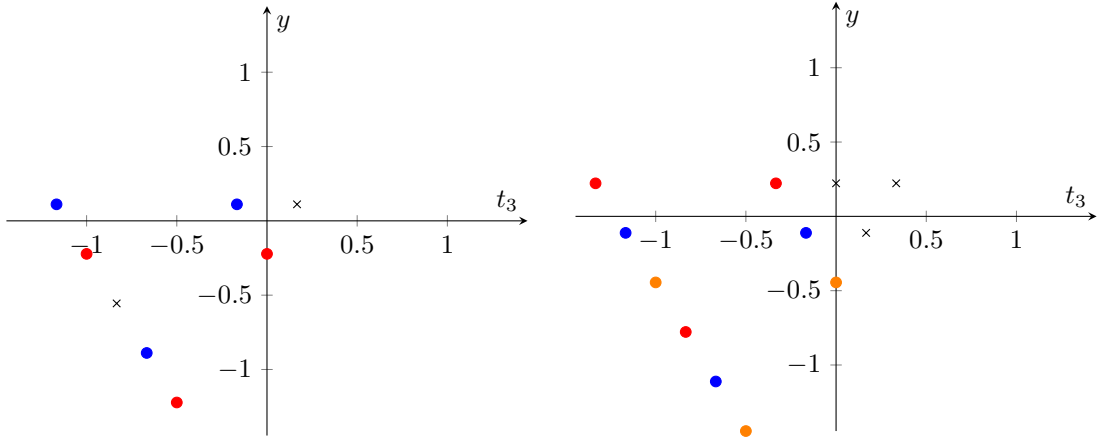
\begin{figure}[h!]
	\begin{center}
		\begin{tikzpicture}
		\begin{axis}[ axis lines=center, grid=none,ymin=-13/9, ymax=13/9, xmin=-13/9, xmax=13/9, xlabel={$t_3$},
		ylabel={$y$}]
		\addplot[
		color=blue,
		mark=*,
		mark options={solid},
		]
		coordinates {
			(-7/6,1/9)
		};
		\addplot[
		color=red,
		mark=*,
		mark options={solid},
		]
		coordinates {
			(-6/6,-2/9)
		};
		\addplot[
		color=black,
		mark=x,
		mark options={solid},
		]
		coordinates {
			(-5/6,-5/9)
		};
		\addplot[
		color=blue,
		mark=*,
		mark options={solid},
		]
		coordinates {
			(-4/6,-8/9)
		};
		\addplot[
		color=red,
		mark=*,
		mark options={solid},
		]
		coordinates {
			(-3/6,-11/9)
		};
		\addplot[
		color=black,
		mark=x,
		mark options={solid},
		]
		coordinates {
			(1/6,1/9)
		};
		\addplot[
		color=blue,
		mark=*,
		mark options={solid},
		]
		coordinates {
			(-1/6,1/9)
		};
		\addplot[
		color=red,
		mark=*,
		mark options={solid},
		]
		coordinates {
			(0,-2/9)
		};
		\end{axis}
		\end{tikzpicture} \hskip 0.5cm
		\begin{tikzpicture}
		\begin{axis}[ axis lines=center, grid=none,ymin=-13/9, ymax=13/9, xmin=-13/9, xmax=13/9, xlabel={$t_3$},
		ylabel={$y$}]
		\addplot[
		color=red,
		mark=*,
		mark options={solid},
		]
		coordinates {
			(-4/3,2/9)
		};
		\addplot[
		color=blue,
		mark=*,
		mark options={solid},
		]
		coordinates {
			(-7/6,-1/9)
		};
		\addplot[
		color=orange,
		mark=*,
		mark options={solid},
		]
		coordinates {
			(-1,-4/9)
		};
		\addplot[
		color=red,
		mark=*,
		mark options={solid},
		]
		coordinates {
			(-5/6,-7/9)
		};
		\addplot[
		color=blue,
		mark=*,
		mark options={solid},
		]
		coordinates {
			(-2/3,-10/9)
		};
		\addplot[
		color=orange,
		mark=*,
		mark options={solid},
		]
		coordinates {
			(-1/2,-13/9)
		};
		
		\addplot[
		color=black,
		mark=x,
		mark options={solid},
		]
		coordinates {
			(1/3,2/9)
		};
		\addplot[
		color=red,
		mark=*,
		mark options={solid},
		]
		coordinates {
			(-1/3,2/9)
		};
		\addplot[
		color=orange,
		mark=*,
		mark options={solid},
		]
		coordinates {
			(0,-4/9)
		};
		\addplot[
		color=black,
		mark=x,
		mark options={solid},
		]
		coordinates {
			(0,2/9)
		};
		\addplot[
		color=black,
		mark=x,
		mark options={solid},
		]
		coordinates {
			(1/6,-1/9)
		};
		\addplot[
		color=blue,
		mark=*,
		mark options={solid},
		]
		coordinates {
			(-1/6,-1/9)
		};
		\end{axis}
		\end{tikzpicture}
	\end{center}
	\caption{Physical states for $N=1$ (left) and $N=2$ (right)  organized in $\mathbf{p}$ multiplets ($m=2$).}
	\label{fig4}
\end{figure}

\noindent When $N=1$, i.e. $\lambda=0$ and $\mu=1$, for $m=2$ eight physical states arise (see \eqref{eq:energydeg}), six of them organized as $\mathbf{3}^2$ and the remaining two as $\mathbf{1}^2$.
For  $N=2$, i.e. $\lambda=0, \mu=2$ twelve states arise, nine of them organized as $\mathbf{3}^3$ and the remaining three as $\mathbf{1}^3$. We represent these situations in \figref{fig4}.  

 \noindent By repeating this procedure for different values of $m$ we have been able to identify the pattern of the multiplets in region II. We report the obtained results in Table \ref{t3}.
\begin{table}[h!]
	\begin{small}
		\begin{center}
			\begin{tabular}{|c c c c c|} 
				\hline
				$\lambda$ & $\mu$ & $\mathbf{p}$ multiplets & Number of Irreps per level & \text{deg}($E_N$)  \\ [0.1ex] 
				\hline
				$0$ & $0,1,2, \dots, m$ & $\begin{cases}  \boldsymbol{3}^{\mu+1}\\ \boldsymbol{1}^{m+\frac{\mu(\mu-1)}{2}} \end{cases}$ & $m+1+\frac{\mu(\mu+1)}{2}$ & $\frac{(\mu+2)(\mu+3)}{2}+m$
				\\ [1ex] 
				\hline
			\end{tabular}
		\end{center}
	\end{small}
\caption{$\mathbf{p}$  multiplets with their number of occurrences, number of irreps of the polynomial algebra and degeneracy per energy level characterizing region II.}
\label{t3}
\end{table}
\newpage \subsubsection*{Region III}
\noindent For $N=m+1,m+2,\dots, 2m+1$, i.e. $\lambda=1$, $\mu=0, 1, 2, \dots, m$ we enter into a new regime, as a new larger $\mathbf{p}$ multiplet arise. This is because we have sufficient room to accomodate connected physical states into a $\mathbf{6}$ multiplet, i.e. we are allowed to ladder two times among physical states with the higher order integrals (but no more than this). 
\noindent The set of physical states in this region, for $\mu=0,1, \dots,m$, can be parametrized as:
\begin{equation}
\ket{N, n_1, n_2} \qquad \text{with} \quad N=\mu+m+1,\quad  n_1=-m-1, 0, \dots, \mu+m+1 \quad \text{and} \quad 0 \leq n_2 \leq \mu+m+1-n_1   \, .
\label{StatesRIII}
\end{equation}
\noindent The are $\#\mathcal{S}(m;1, \mu)=(m+1)(7 m^2+38m+36)/6$ basis states in this region.
Among them, the ones for which:

\begin{equation}
\begin{cases}
n_1+ n_2 =\mu+m+1, \mu+m, \mu+m-1, \dots, \mu+1\\
n_2 =0, 1,\dots, m \, ,
\end{cases}
\label{eq:cases}
\end{equation}

\noindent are all annihilated by the action of the $(\hat{T}_+,\hat{V}_+)$ operators. The are a total number  $\# \tilde{\mathcal{S}}(m;1,\mu)=(m+1)(m+2)(5m+3)$ of these states. Let us take the example $N=m+1$ (i.e. $\lambda=1$, $\mu=0$). We have a total number of $\#\mathcal{S}(m; 1,0)=(m+3)(m+4)/2+m$ physical states:

\begin{table}[h!]
	\begin{small}
		\begin{center}
			$\mathcal{S}(m;1, 0) = \begin{Bmatrix}

\ket{m+1, -m-1,0} & \ket{m+1, 0, 0}  & \ket{m+1,1,0}  & \dots & \ket{m+1, m+1, 0} \\ 
\ket{m+1, -m-1,1} & \ket{m+1, 0, 1}  & \ket{m+1,1,1} & &  \\ 
\vdots &  \vdots & \vdots  &  &   \\
\ket{m+1, -m-1,m} & \ket{m+1, 0, m} & \ket{m+1,1,m}  &  &  \\  
\ket{m+1, -m-1,m+1} & \ket{m+1, 0, m+1} &   &  &   \\  
\vdots & & & &    \\ 
\ket{m+1,-m-1,2m+2} & & & & 
\end{Bmatrix}$
\end{center}
\end{small}
\end{table}
The ones annihilated by $(\hat{T}_+,\hat{V}_+)$ are the ones for which the following system is satisfied:
\begin{equation}
\begin{cases}
n_1+ n_2 =m+1, m, m-1, \dots, 1\\
n_2 =0, 1,\dots, m \, .
\end{cases}
\label{eq:cases0}
\end{equation}

If $n_1=-m-1$, i.e. we consider the first column in $\mathcal{S}(m;1,0)$, we get:

\begin{equation}
\begin{cases}
n_2 =2m+2, 2m+1, 2m-1, \dots, m+2\\
n_2 =0, 1,\dots, m \, .
\end{cases}
\label{eq:cases01}
\end{equation}

Thus, no common solutions exist. For $n_1=0,1,\dots, m+1$, we get:
\begin{equation}
\begin{cases}
n_2 =m+1-n_1, m-n_1, m-1-n_1, \dots, 1-n_1\\
n_2 =0, 1,\dots, m \, ,
\end{cases}
\label{eq:cases03}
\end{equation}

and the solutions are given by:

\begin{equation}
\begin{cases}
n_1 =0 \qquad \quad \rightarrow \quad \quad n_2=1, \dots, m\\
n_1 =1 \qquad\quad \rightarrow \quad \quad n_2=0, \dots, m\\
n_1 =2 \qquad \quad\rightarrow \quad \quad n_2 =0, \dots, m-1\\
n_1 =3 \qquad \quad\rightarrow \quad \quad n_2 =0, \dots, m-2\\
\vdots \hskip 3.2cm \vdots\\
n_1=m+1 \quad \rightarrow \quad \quad n_2 = 0 \, .
\end{cases}
\label{eq:cases02}
\end{equation}

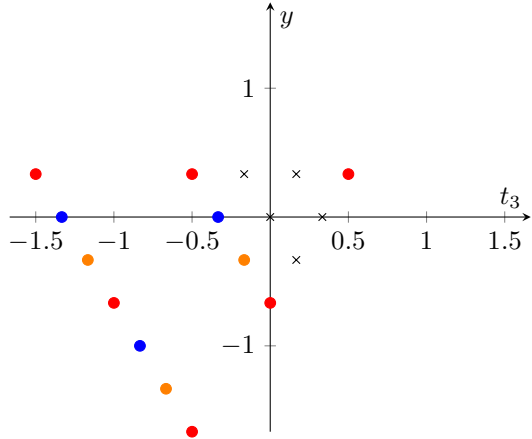
\begin{figure}[httb]
	\begin{center}
		\begin{tikzpicture}
		\begin{axis}[ axis lines=center, grid=none,ymin=-5/3, ymax=5/3, xmin=-5/3, xmax=5/3, xlabel={$t_3$},
		ylabel={$y$}]
		\addplot[
		color=red,
		mark=*,
		mark options={solid},
		]
		coordinates {
			(-3/2,1/3)
		};
		\addplot[
		color=blue,
		mark=*,
		mark options={solid},
		]
		coordinates {
			(-4/3,0)
		};
		\addplot[
		color=orange,
		mark=*,
		mark options={solid},
		]
		coordinates {
			(-7/6,-1/3)
		};
		\addplot[
		color=red,
		mark=*,
		mark options={solid},
		]
		coordinates {
			(-1,-2/3)
		};
		\addplot[
		color=blue,
		mark=*,
		mark options={solid},
		]
		coordinates {
			(-5/6,-1)
		};
		\addplot[
		color=orange,
		mark=*,
		mark options={solid},
		]
		coordinates {
			(-2/3,-4/3)
		};
			\addplot[
		color=red,
		mark=*,
		mark options={solid},
		]
		coordinates {
			(-1/2,-5/3)
		};
		\addplot[
		color=red,
		mark=*,
		mark options={solid},
		]
		coordinates {
			(1/2,1/3)
		};
		\addplot[
		color=red,
		mark=*,
		mark options={solid},
		]
		coordinates {
			(-1/2,1/3)
		};
		\addplot[
		color=red,
		mark=*,
		mark options={solid},
		]
		coordinates {
			(0,-2/3)
		};
		\addplot[
		color=black,
		mark=x,
		mark options={solid},
		]
		coordinates {
			(1/6,1/3)
		};
		\addplot[
		color=black,
		mark=x,
		mark options={solid},
		]
		coordinates {
			(-1/6,1/3)
		};
		\addplot[
		color=black,
		mark=x,
		mark options={solid},
		]
		coordinates {
			(1/6,-1/3)
		};
		\addplot[
	color=black,
	mark=x,
	mark options={solid},
	]
	coordinates {
		(1/3,0)
	};
	\addplot[
color=orange,
mark=*,
mark options={solid},
]
coordinates {
	(-1/6,-1/3)
};
	\addplot[
color=blue,
mark=*,
mark options={solid},
]
coordinates {
	(-1/3,0)
};
	\addplot[
color=black,
mark=x,
mark options={solid},
]
coordinates {
	(0,0)
};
	\end{axis}
	\end{tikzpicture}
	\end{center}
	\caption{Physical states for $N=3$ organized in $\mathbf{p}$ multiplets ($m=2$).}
	\label{fig5}
\end{figure}

\noindent We thus have a total number of $m+(m+1)+(m+(m-1)+\dots +1)=(m+1)(m+2)/2+m$ states annihilated by $\hat{V}_+$ and $\hat{T}_+$. They can be all parametrized as:
\begin{equation}
\ket{m+1, 0, n_2} \quad  n_2=1, \dots, m \quad \cup \quad \ket{m+1, n_1, n_2} \quad \, n_1=1, \dots, m+1 \, , \quad 0 \leq n_2 \leq m-n_1+1 \, .
\label{eq:annbs}
\end{equation}
If we now apply $(\hat{V}_-)^{p+1}$ until we get zero on these states, we can see how the total number of $(m+3)(m+4)/2+m$ states are organized. For example, for $m=2$, the states with $N=3$ turn out to be connected as in \figref{fig5}. 

\noindent Observing \figref{fig5}, we can see that the seventeen states organize as a $\mathbf{6}$ multiplet (red), two $\mathbf{3}$ multiplets (blue, orange), and five singlets under the action of the higher order integrals. 
By repeating  the construction for generic $N=m+1, m+2, \dots, 2m+1$, i.e. $\lambda=1$, $\mu=0,1,\dots, m$ we have been able to identify the pattern of multiplets in region III. We report the obtained results in Table \ref{t4}.
\begin{table}[h!]
	\begin{small}
		\begin{center}
			\begin{tabular}{|c c c c c|} 
					\hline
				$\lambda$ & $\mu$ & $\mathbf{p}$ multiplets & Number of Irreps per level & \text{deg}($E_N$)  \\ [0.1ex] 
				\hline
				$1$&$0, 1, 2, \dots, m$& $\begin{cases}
			\boldsymbol{6}^{\mu+1}\\
			\boldsymbol{3}^{m+\frac{\mu(\mu-1)}{2}}\\
			\boldsymbol{1}^{\frac{m(m+3)}{2} +\mu(m-\mu-1)}\\
			\end{cases}$ &$\footnotesize \frac{(m+1)(m+2)}{2}+\frac{\mu(2m-\mu-1)}{2}+m$& $\frac{(m+\mu+3)(m+\mu+4)}{2}+m$
				\\ [1ex] 
				\hline
			\end{tabular}
		\end{center}
	\end{small}
\caption{$\mathbf{p}$  multiplets with their number of occurrences, number of irreps of the polynomial algebra and degeneracy per energy level characterizing region III.}
\label{t4}
\end{table}

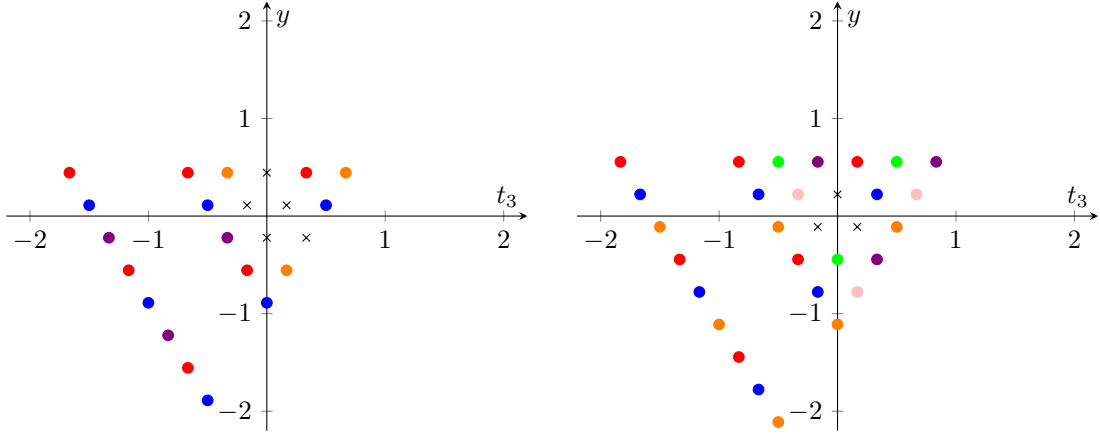
\begin{figure}[httb]
	\begin{center}
		\begin{tikzpicture}
	\begin{axis}[ axis lines=center, grid=none,ymin=-2.2, ymax=2.2, xmin=-2.2, xmax=2.2, xlabel={$t_3$},
	ylabel={$y$}]
		\addplot[
		color=red,
		mark=*,
		mark options={solid},
		]
		coordinates {
		(-10/6,4/9)
		};
	\addplot[
	color=blue,
	mark=*,
	mark options={solid},
	]
	coordinates {
		(-9/6,1/9)
	};
	\addplot[
color=violet,
mark=*,
mark options={solid},
]
coordinates {
	(-8/6,-2/9)
};
	\addplot[
color=red,
mark=*,
mark options={solid},
]
coordinates {
	(-7/6,-5/9)
};
	\addplot[
color=blue,
mark=*,
mark options={solid},
]
coordinates {
	(-6/6,-8/9)
};
	\addplot[
color=violet,
mark=*,
mark options={solid},
]
coordinates {
	(-5/6,-11/9)
};
	\addplot[
color=red,
mark=*,
mark options={solid},
]
coordinates {
	(-4/6,-14/9)
};
	\addplot[
color=blue,
mark=*,
mark options={solid},
]
coordinates {
	(-3/6,-17/9)
};
	\addplot[
color=red,
mark=*,
mark options={solid},
]
coordinates {
	(-4/6,4/9)
};
	\addplot[
color=orange,
mark=*,
mark options={solid},
]
coordinates {
	(4/6,4/9)
};
	\addplot[
color=blue,
mark=*,
mark options={solid},
]
coordinates {
	(0,-8/9)
};
	\addplot[
color=red,
mark=*,
mark options={solid},
]
coordinates {
	(2/6,4/9)
};
	\addplot[
color=blue,
mark=*,
mark options={solid},
]
coordinates {
	(3/6,1/9)
};
	\addplot[
color=black,
mark=x,
mark options={solid},
]
coordinates {
	(0,-2/9)
};
	\addplot[
color=black,
mark=x,
mark options={solid},
]
coordinates {
	(-1/6,1/9)
};
	\addplot[
color=black,
mark=x,
mark options={solid},
]
coordinates {
	(1/6,1/9)
};
	\addplot[
color=orange,
mark=*,
mark options={solid},
]
coordinates {
	(1/6,-5/9)
};
	\addplot[
color=red,
mark=*,
mark options={solid},
]
coordinates {
	(-1/6,-5/9)
};
	\addplot[
color=black,
mark=x,
mark options={solid},
]
coordinates {
	(0,4/9)
};
	\addplot[
color=violet,
mark=*,
mark options={solid},
]
coordinates {
	(-2/6,-2/9)
};
	\addplot[
color=black,
mark=x,
mark options={solid},
]
coordinates {
	(2/6,-2/9)
};
	\addplot[
color=blue,
mark=*,
mark options={solid},
]
coordinates {
	(-3/6,1/9)
};
	\addplot[
color=orange,
mark=*,
mark options={solid},
]
coordinates {
	(-2/6,4/9)
};
		\end{axis}
		\end{tikzpicture} \hskip 0.5cm
		\begin{tikzpicture}
		\begin{axis}[ axis lines=center, grid=none,ymin=-2.2, ymax=2.2, xmin=-2.2, xmax=2.2, xlabel={$t_3$},
		ylabel={$y$}]
		\addplot[
		color=red,
		mark=*,
		mark options={solid},
		]
		coordinates {
			(-11/6,5/9)
		};
		\addplot[
	color=blue,
	mark=*,
	mark options={solid},
	]
	coordinates {
		(-10/6,2/9)
	};
	\addplot[
color=orange,
mark=*,
mark options={solid},
]
coordinates {
	(-9/6,-1/9)
};
	\addplot[
color=red,
mark=*,
mark options={solid},
]
coordinates {
	(-8/6,-4/9)
};
	\addplot[
color=blue,
mark=*,
mark options={solid},
]
coordinates {
	(-7/6,-7/9)
};
	\addplot[
color=orange,
mark=*,
mark options={solid},
]
coordinates {
	(-6/6,-10/9)
};
	\addplot[
color=red,
mark=*,
mark options={solid},
]
coordinates {
	(-5/6,-13/9)
};
	\addplot[
color=blue,
mark=*,
mark options={solid},
]
coordinates {
	(-4/6,-16/9)
};
	\addplot[
color=orange,
mark=*,
mark options={solid},
]
coordinates {
	(-3/6,-19/9)
};
	\addplot[
color=violet,
mark=*,
mark options={solid},
]
coordinates {
	(5/6,5/9)
};
	\addplot[
color=red,
mark=*,
mark options={solid},
]
coordinates {
	(-5/6,5/9)
};
\addplot[
color=orange,
mark=*,
mark options={solid},
]
coordinates {
	(0,-10/9)
};
\addplot[
color=green,
mark=*,
mark options={solid},
]
coordinates {
	(-3/6,5/9)
};
\addplot[
color=red,
mark=*,
mark options={solid},
]
coordinates {
	(1/6,5/9)
};
\addplot[
color=violet,
mark=*,
mark options={solid},
]
coordinates {
	(-1/6,5/9)
};
\addplot[
color=green,
mark=*,
mark options={solid},
]
coordinates {
	(3/6,5/9)
};
\addplot[
color=pink,
mark=*,
mark options={solid},
]
coordinates {
	(-2/6,2/9)
};
\addplot[
color=blue,
mark=*,
mark options={solid},
]
coordinates {
	(-4/6,2/9)
};
\addplot[
color=black,
mark=x,
mark options={solid},
]
coordinates {
	(0,2/9)
};
\addplot[
color=pink,
mark=*,
mark options={solid},
]
coordinates {
	(4/6,2/9)
};
\addplot[
color=blue,
mark=*,
mark options={solid},
]
coordinates {
	(2/6,2/9)
};
\addplot[
color=orange,
mark=*,
mark options={solid},
]
coordinates {
	(3/6,-1/9)
};
\addplot[
color=orange,
mark=*,
mark options={solid},
]
coordinates {
	(-3/6,-1/9)
};
\addplot[
color=black,
mark=x,
mark options={solid},
]
coordinates {
	(1/6,-1/9)
};
\addplot[
color=black,
mark=x,
mark options={solid},
]
coordinates {
	(-1/6,-1/9)
};
\addplot[
color=violet,
mark=*,
mark options={solid},
]
coordinates {
	(2/6,-4/9)
};
\addplot[
color=red,
mark=*,
mark options={solid},
]
coordinates {
	(-2/6,-4/9)
};
\addplot[
color=green,
mark=*,
mark options={solid},
]
coordinates {
	(0,-4/9)
};
\addplot[
color=pink,
mark=*,
mark options={solid},
]
coordinates {
	(1/6,-7/9)
};
\addplot[
color=blue,
mark=*,
mark options={solid},
]
coordinates {
	(-1/6,-7/9)
};
		\end{axis}
		\end{tikzpicture}
	\end{center}
	\caption{Physical states arising for $N=4$ (left)  and $N =5$ (right) organized in $\mathbf{p}$ multiplets  ($m=2$).}
	\label{fig6}
\end{figure}

\noindent For $m=2$, states with $N=4,5$ also belong to this region. We report the multiplets in \figref{fig6}.
\newpage
\subsubsection*{Region IV}
\noindent If we go ahead in the construction by considering higher values of $N$, we realize that other regimes arise for $N=2m+2, 2m+3, \dots, 3m+2$, $N=3m+3,3m+4, \dots, 4m+2$ etc. In particular, for $\lambda=2,3 \dots$ and $\mu =0,1, \dots, m$, the set of basis states is the following:
\begin{equation*}
\ket{N, n_1, n_2} \qquad \text{with} \quad N=(m+1)\lambda+\mu,\quad  n_1=-m-1, 0, \dots, (m+1)\lambda+\mu \quad \text{and} \quad 0 \leq n_2 \leq (m+1)\lambda+\mu-n_1 \, .
\end{equation*}
Let us focus on the first sub-region $\lambda=2$, $\mu=0,1,2,\dots,m$. We have:

	\begin{small}
		\begin{center}
			$\mathcal{S}(m;2, \mu)=$ \vskip 0.4cm
			
			 $\begin{Bmatrix}
				\ket{2m+2, -m-1,0} & \ket{2m+2, 0, 0}  & \ket{2m+2,1,0}  & \dots & \ket{2m+2, 2m+1,0}&\ket{2m+2, 2m+2, 0} \\ 
				\ket{2m+2, -m-1,1} & \ket{2m+2, 0, 1}  & \ket{2m+2,1,1} &\dots & \ket{2m+2,2m+1,1}&  \\ 
				\vdots &  \vdots& \vdots  &  & &  \\
				\ket{2m+2, -m-1,2m+1} & \ket{2m+2, 0, 2m+1} & \ket{2m+2,1,2m+1} &  & &  \\  
				 \ket{2m+2, -m-1,2m+2} & \ket{2m+2, 0, 2m+2} &   &  &  & \\  
				\vdots & & & &    &\\ 
				\ket{2m+2,-m-1,3m+3} & & & & &
			\end{Bmatrix}$ \vskip 0.4 cm $\cup \dots \cup$ \vskip 0.4 cm 
		
		 $\begin{Bmatrix}
			\ket{3m+2, -m-1,0} &\ket{3m+2, 0, 0}  & \ket{3m+2,1,0}  & \dots & \ket{3m+2, 3m+1,0} & \ket{3m+2, 3m+2, 0} \\ 
			\ket{3m+2, -m-1,1} & \ket{3m+2, 0, 1}  & \ket{3m+2,1,1} & \dots  & \ket{3m+2,3m+1,1}&  \\ 
			\vdots &  \vdots & \vdots   &  & &  \\
			\ket{3m+2, -m-1,3m+1} & \ket{3m+2, 0, 3m+1} & \ket{3m+2,1,3m+1}  &  & &  \\  
			\ket{3m+2, -m-1,3m+2} & \ket{3m+2, 0, 3m+2} &   &  &  & \\  
			\vdots & & & &    &\\ 
			\ket{3m+2,-m-1,4m+3} & & & & &
		\end{Bmatrix}$ 
		\end{center}
	\end{small}

\noindent Again, by imposing: $$(\hat{T}_+,\hat{V}_+) \rhd \{\mathcal{S}(m;2,0),\mathcal{S}(m;2,1), \dots, \mathcal{S}(m;2,m)\}=0 \, ,$$ 

we select the subset $\tilde{\mathcal{S}}(m;2,\mu)$ composed by $\# \tilde{\mathcal{S}}(m;2,\mu)=(m+1)^3$ states annihilated by the above operators. Once these states are found, we act with $(\hat{V}_-)^{p+1}$ to specify the boundary of the multiplets of connected states. 

Let us perform the construction for $N=2m+2$, i.e. $\lambda=2, \mu=0$ to provide some explicit example inside this region. For $\lambda=2, \mu=0$ the following set of $\#\mathcal{S}(m; 2,0)=(2m+4)(2m+5)/2+m$ physical states arise:

\begin{small}
\begin{center}
				 $\mathcal{S}(m;2, 0)$ = \vskip 0.4cm$\begin{Bmatrix}
	
	\ket{2m+2, -m-1,0} & \ket{2m+2, 0, 0}  & \ket{2m+2,1,0}  & \dots & \ket{2m+2, 2m+1,0} & \ket{2m+2, 2m+2, 0} \\ 
	\ket{2m+2, -m-1,1} & \ket{2m+2, 0, 1}  & \ket{2m+2,1,1} &\dots & \ket{2m+2,2m+1,1}&  \\ 
	\vdots &  \vdots & \vdots  &  & &  \\
	\ket{2m+2, -m-1,2m+1} & \ket{2m+2, 0, 2m+1} & \ket{2m+2,1,2m+1}  &  & &  \\  
	\ket{2m+2, -m-1,2m+2} & \ket{2m+2, 0, 2m+2} &   &  &  & \\  
	\vdots & & & &    &\\ 
	\ket{2m+2,-m-1,3m+3} & & & & &
\end{Bmatrix}$ 
\end{center}
\end{small}

\noindent The ones annihilated by $(\hat{T}_+,\hat{V}_+)$, i.e. the subset $\tilde{\mathcal{S}}(m;2,0)$, are those for which the following system is satisfied:
\begin{equation}
\begin{cases}
n_1+ n_2 =2m+2, 2m+1, \dots, m+3, m+2\\
n_2 =0, 1,\dots, m \, ,
\end{cases}
\label{eq:cases20}
\end{equation}

\noindent namely:

\begin{equation}
\begin{cases}
n_1 =2 \qquad \quad \,\, \rightarrow \quad \quad n_2=m\\
n_1 =3 \qquad \quad\,\,  \rightarrow \quad \quad n_2=m-1,  m\\
\vdots \hskip 3.3cm \vdots \\
n_1=m \qquad  \,\,\,\,\,\,\rightarrow \quad\quad  n_2 = 2,\dots,m \\
n_1=m+1\quad \,\,\rightarrow \quad \quad n_2 = 1,\dots,m\\
n_1=m+2\quad \,\,\rightarrow \quad \quad n_2 = 0,\dots,m\\
\vdots \hskip 3.3cm \vdots\\
n_1 =2m+1 \quad \rightarrow \quad \quad n_2 = 0,1 \\
n_1 =2m+2 \quad \rightarrow \quad \quad n_2 = 0 \, .
\end{cases}
\label{eq:cas}
\end{equation}

\noindent We thus have a total number of $(1+2+\dots +m+1)+(m+(m-1)+\dots +1)=(m+1)(m+2)/2+m(m+1)/2=(m+1)^2$ states annihilated by the action of both $\hat{V}_+$ and $\hat{T}_+$. Thus, $\#\tilde{\mathcal{S}}(m;2,0)=(m+1)^2$. They can be  parametrized as:
$$\ket{2m+2, n_1, n_2} \quad  n_1=2, \dots, m+2 \, , \quad m+2-n_1 \leq n_2 \leq m \quad \cup \quad    n_1=m+3, \dots, 2m+2 \, , \quad 0 \leq n_2 \leq 2m+2-n_1 \, .$$
If we now apply $(\hat{V}_-)^{p+1}$ until we get zero on these states, we can see how the total number of $(2m+4)(2m+5)/2+m$ states are organized under the action of the higher order integrals. The result we get is that they decompose as $\mathbf{10}$, $\mathbf{6}^m$, $\mathbf{3}^{m(m+3)/2}$ and $\mathbf{1}^{m(m-1)/2}$. Clearly, $10+6m+3m(m+3)/2+m(m-1)/2=(2m+4)(2m+5)+m$. Now, by repeating this process for each $\lambda=3,4,\dots$, for $\mu=0,1,\dots, m$ we have been able to identify the patterns for all the multiplets. 

 We report the results that we have obtained in Table \ref{t5}. 

\begin{figure}[httb]
	\begin{center}
		\begin{tikzpicture}
	\begin{axis}[ axis lines=center, grid=none,ymin=-25/9, ymax=25/9, xmin=-25/9, xmax=25/9, xlabel={$t_3$},
ylabel={$y$}]
		\addplot[
		color=orange,
		mark=*,
		mark options={solid},
		]
		coordinates {
			(-12/6,6/9)
		};
		\addplot[
		color=blue,
		mark=*,
		mark options={solid},
		]
		coordinates {
			(-11/6,3/9)
		};
		\addplot[
		color=gray,
		mark=*,
		mark options={solid},
		]
		coordinates {
			(-10/6,0)
		};
		\addplot[
		color=orange,
		mark=*,
		mark options={solid},
		]
		coordinates {
			(-9/6,-3/9)
		};
		\addplot[
		color=blue,
		mark=*,
		mark options={solid},
		]
		coordinates {
			(-8/6,-6/9)
		};
		\addplot[
		color=gray,
		mark=*,
		mark options={solid},
		]
		coordinates {
			(-7/6,-9/9)
		};
		\addplot[
		color=orange,
		mark=*,
		mark options={solid},
		]
		coordinates {
			(-6/6,-12/9)
		};
		\addplot[
		color=blue,
		mark=*,
		mark options={solid},
		]
		coordinates {
			(-5/6,-15/9)
		};
		\addplot[
		color=gray,
		mark=*,
		mark options={solid},
		]
		coordinates {
			(-4/6,-18/9)
		};
		\addplot[
		color=orange,
		mark=*,
		mark options={solid},
		]
		coordinates {
			(-3/6,-21/9)
		};
		\addplot[
		color=orange,
		mark=*,
		mark options={solid},
		]
		coordinates {
			(6/6,6/9)
		};
	\addplot[
	color=red,
	mark=*,
	mark options={solid},
	]
	coordinates {
		(4/6,6/9)
	};
	\addplot[
color=violet,
mark=*,
mark options={solid},
]
coordinates {
	(2/6,6/9)
};
	\addplot[
color=orange,
mark=*,
mark options={solid},
]
coordinates {
	(0,6/9)
};
	\addplot[
color=red,
mark=*,
mark options={solid},
]
coordinates {
	(-2/6,6/9)
};
	\addplot[
color=violet,
mark=*,
mark options={solid},
]
coordinates {
	(-4/6,6/9)
};
	\addplot[
color=orange,
mark=*,
mark options={solid},
]
coordinates {
	(-6/6,6/9)
};
	\addplot[
color=blue,
mark=*,
mark options={solid},
]
coordinates {
	(-5/6,3/9)
};
	\addplot[
color=yellow,
mark=*,
mark options={solid},
]
coordinates {
	(-3/6,3/9)
};
	\addplot[
color=purple,
mark=*,
mark options={solid},
]
coordinates {
	(-1/6,3/9)
};
	\addplot[
color=blue,
mark=*,
mark options={solid},
]
coordinates {
	(1/6,3/9)
};
	\addplot[
color=yellow,
mark=*,
mark options={solid},
]
coordinates {
	(3/6,3/9)
};
	\addplot[
color=purple,
mark=*,
mark options={solid},
]
coordinates {
	(5/6,3/9)
};
	\addplot[
color=gray,
mark=*,
mark options={solid},
]
coordinates {
	(-4/6,0)
};
	\addplot[
color=green,
mark=*,
mark options={solid},
]
coordinates {
	(-2/6,0)
};
	\addplot[
color=black,
mark=x,
mark options={solid},
]
coordinates {
	(0,0)
};
	\addplot[
color=gray,
mark=*,
mark options={solid},
]
coordinates {
	(2/6,0)
};
	\addplot[
color=green,
mark=*,
mark options={solid},
]
coordinates {
	(4/6,0)
};
	\addplot[
color=orange,
mark=*,
mark options={solid},
]
coordinates {
	(-3/6,-3/9)
};
	\addplot[
color=violet,
mark=*,
mark options={solid},
]
coordinates {
	(-1/6,-3/9)
};
	\addplot[
color=red,
mark=*,
mark options={solid},
]
coordinates {
	(1/6,-3/9)
};
	\addplot[
color=orange,
mark=*,
mark options={solid},
]
coordinates {
	(3/6,-3/9)
};
	\addplot[
color=blue,
mark=*,
mark options={solid},
]
coordinates {
	(-2/6,-6/9)
};
	\addplot[
color=yellow,
mark=*,
mark options={solid},
]
coordinates {
	(0,-6/9)
};
	\addplot[
color=purple,
mark=*,
mark options={solid},
]
coordinates {
	(2/6,-6/9)
};
	\addplot[
color=gray,
mark=*,
mark options={solid},
]
coordinates {
	(-1/6,-9/9)
};
	\addplot[
color=green,
mark=*,
mark options={solid},
]
coordinates {
	(1/6,-9/9)
};
	\addplot[
color=orange,
mark=*,
mark options={solid},
]
coordinates {
	(0,-12/9)
};

		\end{axis}
		\end{tikzpicture} \hskip 0.5cm
				\begin{tikzpicture}
	\begin{axis}[ axis lines=center, grid=none,ymin=-25/9, ymax=25/9, xmin=-25/9, xmax=25/9, xlabel={$t_3$},
ylabel={$y$}]
		\addplot[
		color=orange,
		mark=*,
		mark options={solid},
		]
		coordinates {
			(-13/6,7/9)
		};
		\addplot[
		color=blue,
		mark=*,
		mark options={solid},
		]
		coordinates {
			(-12/6,4/9)
		};
		\addplot[
		color=gray,
		mark=*,
		mark options={solid},
		]
		coordinates {
			(-11/6,1/9)
		};
		\addplot[
		color=orange,
		mark=*,
		mark options={solid},
		]
		coordinates {
			(-10/6,-2/9)
		};
		\addplot[
		color=blue,
		mark=*,
		mark options={solid},
		]
		coordinates {
			(-9/6,-5/9)
		};
		\addplot[
		color=gray,
		mark=*,
		mark options={solid},
		]
		coordinates {
			(-8/6,-8/9)
		};
		\addplot[
		color=orange,
		mark=*,
		mark options={solid},
		]
		coordinates {
			(-7/6,-11/9)
		};
		\addplot[
		color=blue,
		mark=*,
		mark options={solid},
		]
		coordinates {
			(-6/6,-14/9)
		};
		\addplot[
		color=gray,
		mark=*,
		mark options={solid},
		]
		coordinates {
			(-5/6,-17/9)
		};
		\addplot[
		color=orange,
		mark=*,
		mark options={solid},
		]
		coordinates {
			(-4/6,-20/9)
		};
		\addplot[
		color=blue,
		mark=*,
		mark options={solid},
		]
		coordinates {
			(-3/6,-23/9)
		};
		\addplot[
		color=red,
		mark=*,
		mark options={solid},
		]
		coordinates {
			(7/6,7/9)
		};
		\addplot[
		color=orange,
		mark=*,
		mark options={solid},
		]
		coordinates {
			(-7/6,7/9)
		};
		\addplot[
		color=blue,
		mark=*,
		mark options={solid},
		]
		coordinates {
			(0,-14/9)
		};
		\addplot[
		color=orange,
		mark=*,
		mark options={solid},
		]
		coordinates {
			(5/6,7/9)
		};
		\addplot[
		color=blue,
		mark=*,
		mark options={solid},
		]
		coordinates {
			(6/6,4/9)
		};
		\addplot[
		color=red,
		mark=*,
		mark options={solid},
		]
		coordinates {
			(-5/6,7/9)
		};
		\addplot[
		color=blue,
		mark=*,
		mark options={solid},
		]
		coordinates {
			(-6/6,4/9)
		};
		\addplot[
		color=red,
		mark=*,
		mark options={solid},
		]
		coordinates {
			(1/6,-11/9)
		};
		\addplot[
		color=orange,
		mark=*,
		mark options={solid},
		]
		coordinates {
			(-1/6,-11/9)
		};
		\addplot[
		color=violet,
		mark=*,
		mark options={solid},
		]
		coordinates {
			(3/6,7/9)
		};
		\addplot[
		color=green,
		mark=*,
		mark options={solid},
		]
		coordinates {
			(5/6,1/9)
		};
		
		\addplot[
		color=violet,
		mark=*,
		mark options={solid},
		]
		coordinates {
			(-3/6,7/9)
		};
		\addplot[
		color=gray,
		mark=*,
		mark options={solid},
		]
		coordinates {
			(-5/6,1/9)
		};
		
		\addplot[
		color=green,
		mark=*,
		mark options={solid},
		]
		coordinates {
			(2/6,-8/9)
		};
		\addplot[
		color=gray,
		mark=*,
		mark options={solid},
		]
		coordinates {
			(-2/6,-8/9)
		};
		
		\addplot[
		color=red,
		mark=*,
		mark options={solid},
		]
		coordinates {
			(1/6,7/9)
		};
		\addplot[
		color=orange,
		mark=*,
		mark options={solid},
		]
		coordinates {
			(-4/6,-2/9)
		};
		
		\addplot[
		color=red,
		mark=*,
		mark options={solid},
		]
		coordinates {
			(4/6,-2/9)
		};
		\addplot[
		color=orange,
		mark=*,
		mark options={solid},
		]
		coordinates {
			(-1/6,7/9)
		};
		
		\addplot[
		color=blue,
		mark=*,
		mark options={solid},
		]
		coordinates {
			(-3/6,-5/9)
		};
		\addplot[
		color=blue,
		mark=*,
		mark options={solid},
		]
		coordinates {
			(3/6,-5/9)
		};
		
		\addplot[
		color=purple,
		mark=*,
		mark options={solid},
		]
		coordinates {
			(0,-8/9)
		};
		\addplot[
		color=pink,
		mark=*,
		mark options={solid},
		]
		coordinates {
			(-4/6,4/9)
		};
		\addplot[
		color=yellow,
		mark=*,
		mark options={solid},
		]
		coordinates {
			(4/6,4/9)
		};
		
		\addplot[
		color=violet,
		mark=*,
		mark options={solid},
		]
		coordinates {
			(0,-2/9)
		};
		
		\addplot[
		color=green,
		mark=*,
		mark options={solid},
		]
		coordinates {
			(-1/6,1/9)
		};
		
		\addplot[
		color=gray,
		mark=*,
		mark options={solid},
		]
		coordinates {
			(1/6,1/9)
		};
		
		\addplot[
		color=orange,
		mark=*,
		mark options={solid},
		]
		coordinates {
			(2/6,-2/9)
		};
		\addplot[
		color=red,
		mark=*,
		mark options={solid},
		]
		coordinates {
			(-2/6,-2/9)
		};
		
		\addplot[
		color=pink,
		mark=*,
		mark options={solid},
		]
		coordinates {
			(-1/6,-5/9)
		};
		\addplot[
		color=yellow,
		mark=*,
		mark options={solid},
		]
		coordinates {
			(1/6,-5/9)
		};
		
		\addplot[
		color=purple,
		mark=*,
		mark options={solid},
		]
		coordinates {
			(3/6,1/9)
		};
		\addplot[
		color=purple,
		mark=*,
		mark options={solid},
		]
		coordinates {
			(-3/6,1/9)
		};
		
		\addplot[
		color=pink,
		mark=*,
		mark options={solid},
		]
		coordinates {
			(2/6,4/9)
		};
		\addplot[
		color=yellow,
		mark=*,
		mark options={solid},
		]
		coordinates {
			(-2/6,4/9)
		};
		
		\addplot[
		color=blue,
		mark=*,
		mark options={solid},
		]
		coordinates {
			(0,4/9)
		};
		
		\end{axis}
		\end{tikzpicture} 
				\begin{tikzpicture}
	\begin{axis}[ axis lines=center, grid=none,ymin=-25/9, ymax=25/9, xmin=-25/9, xmax=25/9, xlabel={$t_3$},
	ylabel={$y$}]
		\addplot[
		color=orange,
		mark=*,
		mark options={solid},
		]
		coordinates {
			(-14/6,8/9)
		};
		\addplot[
		color=blue,
		mark=*,
		mark options={solid},
		]
		coordinates {
			(-13/6,5/9)
		};
		\addplot[
		color=gray,
		mark=*,
		mark options={solid},
		]
		coordinates {
			(-12/6,2/9)
		};
		\addplot[
		color=orange,
		mark=*,
		mark options={solid},
		]
		coordinates {
			(-11/6,-1/9)
		};
		\addplot[
		color=blue,
		mark=*,
		mark options={solid},
		]
		coordinates {
			(-10/6,-4/9)
		};
		\addplot[
		color=gray,
		mark=*,
		mark options={solid},
		]
		coordinates {
			(-9/6,-7/9)
		};
		\addplot[
		color=orange,
		mark=*,
		mark options={solid},
		]
		coordinates {
			(-8/6,-10/9)
		};
		\addplot[
		color=blue,
		mark=*,
		mark options={solid},
		]
		coordinates {
			(-7/6,-13/9)
		};
		\addplot[
		color=gray,
		mark=*,
		mark options={solid},
		]
		coordinates {
			(-6/6,-16/9)
		};
		\addplot[
		color=orange,
		mark=*,
		mark options={solid},
		]
		coordinates {
			(-5/6,-19/9)
		};
		\addplot[
		color=blue,
		mark=*,
		mark options={solid},
		]
		coordinates {
			(-4/6,-22/9)
		};
		\addplot[
	color=gray,
	mark=*,
	mark options={solid},
	]
	coordinates {
		(-3/6,-25/9)
	};
		\addplot[
	color=orange,
	mark=*,
	mark options={solid},
	]
	coordinates {
		(-8/6, 8/9)
	};
	\addplot[
color=violet,
mark=*,
mark options={solid},
]
coordinates {
	(-6/6,8/9)
};
	\addplot[
color=red,
mark=*,
mark options={solid},
]
coordinates {
	(-4/6,8/9)
};
	\addplot[
color=orange,
mark=*,
mark options={solid},
]
coordinates {
	(-2/6,8/9)
};
	\addplot[
color=violet,
mark=*,
mark options={solid},
]
coordinates {
	(0,8/9)
};
	\addplot[
color=red,
mark=*,
mark options={solid},
]
coordinates {
	(2/6,8/9)
};
	\addplot[
color=orange,
mark=*,
mark options={solid},
]
coordinates {
	(4/6,8/9)
};
	\addplot[
color=violet,
mark=*,
mark options={solid},
]
coordinates {
	(6/6,8/9)
};
	\addplot[
color=red,
mark=*,
mark options={solid},
]
coordinates {
	(8/6,8/9)
};

	\addplot[
color=blue,
mark=*,
mark options={solid},
]
coordinates {
	(-7/6,5/9)
};
	\addplot[
color=green,
mark=*,
mark options={solid},
]
coordinates {
	(-5/6,5/9)
};
	\addplot[
color=yellow,
mark=*,
mark options={solid},
]
coordinates {
	(-3/6,5/9)
};
	\addplot[
color=blue,
mark=*,
mark options={solid},
]
coordinates {
	(-1/6,5/9)
};
	\addplot[
color=green,
mark=*,
mark options={solid},
]
coordinates {
	(1/6,5/9)
};
	\addplot[
color=yellow,
mark=*,
mark options={solid},
]
coordinates {
	(3/6,5/9)
};
	\addplot[
color=blue,
mark=*,
mark options={solid},
]
coordinates {
	(5/6,5/9)
};
	\addplot[
color=green,
mark=*,
mark options={solid},
]
coordinates {
	(7/6,5/9)
};

	\addplot[
color=gray,
mark=*,
mark options={solid},
]
coordinates {
	(-6/6,2/9)
};
	\addplot[
color=purple,
mark=*,
mark options={solid},
]
coordinates {
	(-4/6,2/9)
};
	\addplot[
color=pink,
mark=*,
mark options={solid},
]
coordinates {
	(-2/6,2/9)
};
	\addplot[
color=gray,
mark=*,
mark options={solid},
]
coordinates {
	(0,2/9)
};
	\addplot[
color=purple,
mark=*,
mark options={solid},
]
coordinates {
	(2/6,2/9)
};
	\addplot[
color=pink,
mark=*,
mark options={solid},
]
coordinates {
	(4/6,2/9)
};
	\addplot[
color=gray,
mark=*,
mark options={solid},
]
coordinates {
	(6/6,2/9)
};

	\addplot[
color=orange,
mark=*,
mark options={solid},
]
coordinates {
	(-5/6,-1/9)
};

	\addplot[
color=violet,
mark=*,
mark options={solid},
]
coordinates {
	(-3/6,-1/9)
};
	\addplot[
color=red,
mark=*,
mark options={solid},
]
coordinates {
	(-1/6,-1/9)
};
	\addplot[
color=orange,
mark=*,
mark options={solid},
]
coordinates {
	(1/6,-1/9)
};
	\addplot[
color=violet,
mark=*,
mark options={solid},
]
coordinates {
	(3/6,-1/9)
};
	\addplot[
color=red,
mark=*,
mark options={solid},
]
coordinates {
	(5/6,-1/9)
};

	\addplot[
color=blue,
mark=*,
mark options={solid},
]
coordinates {
	(-4/6,-4/9)
};
	\addplot[
color=green,
mark=*,
mark options={solid},
]
coordinates {
	(-2/6,-4/9)
};
	\addplot[
color=yellow,
mark=*,
mark options={solid},
]
coordinates {
	(0,-4/9)
};
	\addplot[
color=blue,
mark=*,
mark options={solid},
]
coordinates {
	(2/6,-4/9)
};
	\addplot[
color=green,
mark=*,
mark options={solid},
]
coordinates {
	(4/6,-4/9)
};

	\addplot[
color=gray,
mark=*,
mark options={solid},
]
coordinates {
	(-3/6,-7/9)
};

	\addplot[
color=purple,
mark=*,
mark options={solid},
]
coordinates {
	(-1/6,-7/9)
};
	\addplot[
color=pink,
mark=*,
mark options={solid},
]
coordinates {
	(1/6,-7/9)
};
	\addplot[
color=gray,
mark=*,
mark options={solid},
]
coordinates {
	(3/6,-7/9)
};

	\addplot[
color=orange,
mark=*,
mark options={solid},
]
coordinates {
	(-2/6,-10/9)
};
	\addplot[
color=violet,
mark=*,
mark options={solid},
]
coordinates {
	(0,-10/9)
};
	\addplot[
color=red,
mark=*,
mark options={solid},
]
coordinates {
	(2/6,-10/9)
};
	\addplot[
color=blue,
mark=*,
mark options={solid},
]
coordinates {
	(-1/6,-13/9)
};
	\addplot[
color=green,
mark=*,
mark options={solid},
]
coordinates {
	(1/6,-13/9)
};
	\addplot[
color=gray,
mark=*,
mark options={solid},
]
coordinates {
	(0,-16/9)
};

		\end{axis}
		\end{tikzpicture}
		
	\end{center}
	\caption{Physical states for $N=6$ (upper left), $N=7$ (upper right) and $N=8$ organized in $\mathbf{p}$ multiplets ($m=2$).}
	\label{fig7}
\end{figure}
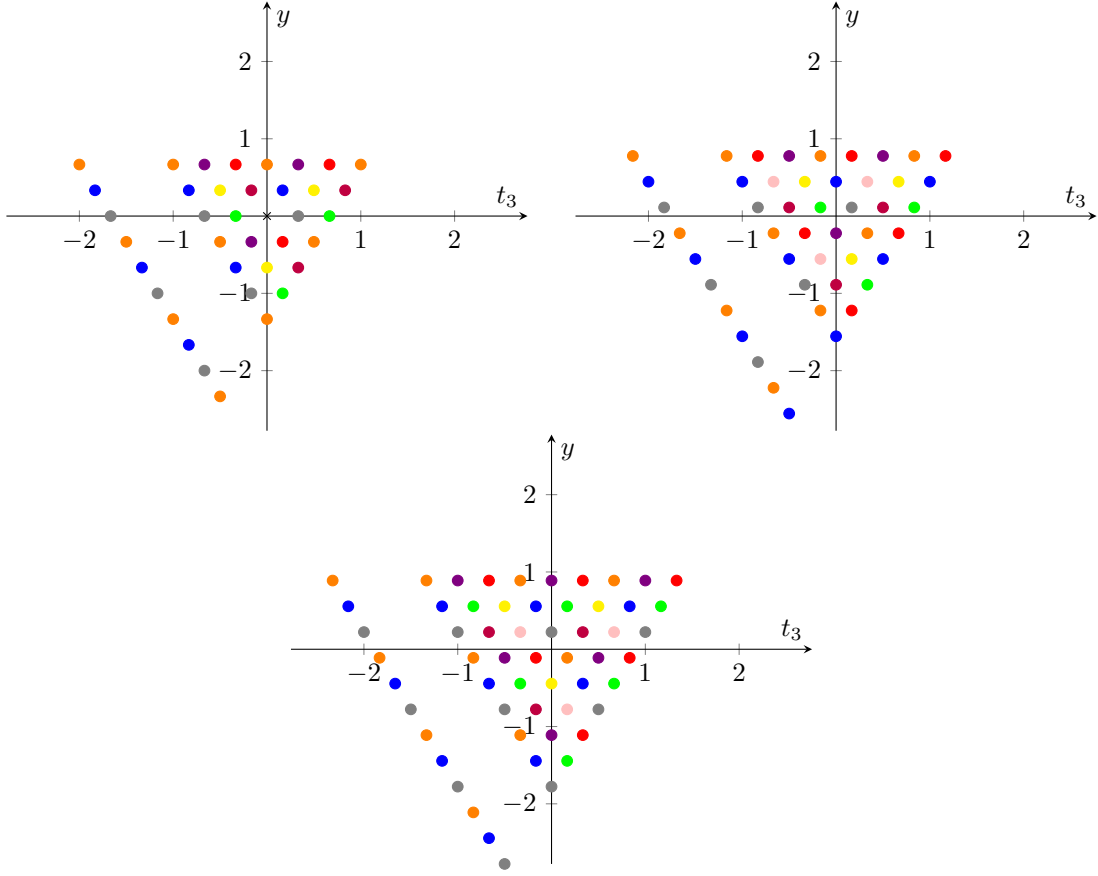

\begin{table}[h!]
	\begin{small}
			\begin{center}
		\begin{tabular}{|c c c c c|} 
				\hline
			$\lambda$ & $\mu$ & $\mathbf{p}$ multiplets& Number of Irreps per level & \text{deg}($E_N$)  \\ [0.1ex] 
			\hline
			$2,3, \dots$ & $0,1,2,\dots, m$ & $\begin{cases}
			\boldsymbol{\frac{(\lambda+2)(\lambda+3)}{2} }^{\mu+1}\\
			\boldsymbol{\frac{(\lambda+1)(\lambda+2)}{2}}^{m+\frac{\mu(\mu-1)}{2}}\\
			\boldsymbol{\frac{(\lambda)(\lambda+1)}{2}}^{\frac{m(m+3)}{2}+\mu(m-\mu-1)}\\
			\boldsymbol{\frac{(\lambda-1)(\lambda)}{2}}^{\frac{(m-\mu)(m-\mu-1)}{2}}
			\end{cases}$  & $(m+1)^2$ & $\frac{((m+1) \lambda+\mu+2)((m+1) \lambda+\mu+3)}{2}+m$  \\ [1ex] 
			\hline
		\end{tabular}
	\end{center}
	\end{small}
\caption{$\mathbf{p}$  multiplets with their number of occurrences, number of irreps and degeneracy per energy level.}
\label{t5}
\end{table}
\noindent Let us fix, just as a clarifying example, $m=2$ and consider the case $\lambda=2, \mu=0,1,2$ i.e. $N=6,7,8$.
For $N=6$, $\#\mathcal{S}(2;2,0)=38$ states $\mathcal{S}(2;2,0)$ arise. Among them, there are $\#\tilde{\mathcal{S}}(2;2,0)=9$ states belonging to the subset $\tilde{\mathcal{S}}(2;2,0)$. For $N=7$, $\#\mathcal{S}(2;2,1)=47$ states $\mathcal{S}(2;2,0)$ arise. Among them, there are $\#\tilde{\mathcal{S}}(2;2,1)=9$ states belonging to the subset $\tilde{\mathcal{S}}(2;2,1)$.  Finally, for $N=8$, $\#\mathcal{S}(2;2,2)=57$ states $\mathcal{S}(2;2,2)$ arise. Among them, there are again $\#\tilde{\mathcal{S}}(2;2,2)=9$ states belonging to the subset $\tilde{\mathcal{S}}(2;2,2)$.  These states, and the way they are connected through the action of the higher order integrals is reported in \figref{fig7}.

\noindent To summarize, for each energy level all multiplets composed by  physical states connected through the action of the higher order integrals have been completely classified for all of the four regions into we splitted up our initial problem. The general results we have obtained are reported in Table \ref{t6}.

\begin{table}[h!]
	\begin{small}
	\begin{center}
		\begin{tabular}{|c| c |c |c |c |c|} 
			\hline
		&	$\lambda$ & $\mu$ & $\mathbf{p}$ multiplets& Number of Irreps per level & \text{deg}($E_N$)  \\
			\hline
		$\textbf{I}$ & $-1$ & $0,1,2,\dots,m$ & $\boldsymbol{1}^{\mu+1}$ & $\mu+1$ & $\mu+1$ \\ 
			\hline
		$\textbf{II}$	&	$0$ & $0,1,2, \dots, m$ & $\begin{cases}  \boldsymbol{3}^{\mu+1}\\ 	
			\boldsymbol{1}^{m+\frac{\mu(\mu-1)}{2}} \end{cases}$ & $m+1+\frac{\mu(\mu+1)}{2}$ & $\frac{(\mu+2)(\mu+3)}{2}+m$\\
			\hline
			$\textbf{III}	$&	$1$ & $0, 1, 2, \dots, m$& $\begin{cases}
			\boldsymbol{6}^{\mu+1}\\
			\boldsymbol{3}^{m+\frac{\mu(\mu-1)}{2}}\\
			\boldsymbol{1}^{\frac{m(m+3)}{2} +\mu(m-\mu-1)}\\
			\end{cases}$ &$\footnotesize \frac{(m+1)(m+2)}{2}+\frac{\mu(2m-\mu-1)}{2}+m$& $\frac{(m+\mu+3)(m+\mu+4)}{2}+m$\\
			\hline
		$	\textbf{IV}$	&	$2,3, \dots$ & $0,1,2,\dots, m$ & $\begin{cases}
			\boldsymbol{\frac{(\lambda+2)(\lambda+3)}{2} }^{\mu+1}\\
			\boldsymbol{\frac{(\lambda+1)(\lambda+2)}{2}}^{m+\frac{\mu(\mu-1)}{2}}\\
			\boldsymbol{\frac{(\lambda)(\lambda+1)}{2}}^{\frac{m(m+3)}{2}+\mu(m-\mu-1)}\\
			\boldsymbol{\frac{(\lambda-1)(\lambda)}{2}}^{\frac{(m-\mu)(m-\mu-1)}{2}}
			\end{cases}$  & $(m+1)^2$ & $\frac{((m+1) \lambda+\mu+2)((m+1) \lambda+\mu+3)}{2}+m$  \\ [1ex] 
			\hline
		\end{tabular}
	\end{center}
	\end{small}
\caption{$\mathbf{p}$  multiplets with their number of occurrences, number of irreps of the polynomial algebra and degeneracy per energy level for the 3D rationally extended harmonic oscillator \eqref{eq:rationallyextended}.}
\label{t6}
\end{table}

\newpage

\section{Concluding Remarks}
\label{conrem}

\noindent In this work we have investigated the polynomial algebras associated with $n$-dimensional superintegrable systems, with separation of variables in Cartesian coordinates, that are endowed with higher order ladder operators satisfying polynomial deformations of the Heisenberg algebra. These models are part of a very large class of superintegrable systems (in fact the classification of one-dimensional systems with operator algebra is still an ongoing problem itself \cite{Marquette2019}). They have connection to Painlevé transcendents \cite{Marquette2009c, Marquette2019a, Marquette2019b} but in general they are associated with nonlinear equations of Painlevé type. The study of these equations is itself a difficult problem as shown by a series of papers by Cosgrove \cite{Cosgrove1,Cosgrove2}. However, there are families of deformations that are of particular interest and that can be classified completely. Among them is the case of $k$-step extension of the harmonic oscillator \cite{ Marquette2014}, which is related to the multi-indexed Hermite exceptional orthogonal polynomials. 

We have introduced a new three-dimensional model involving Hermite exceptional orthogonal polynomials of type III. This system exhibits very interesting degeneracies for its energy spectrum. One of the key results is the application of the underlying  polynomial algebra structure to characterize degeneracies in terms of its finite-dimensional representations. We have performed calculations in order to obtain explicit expressions for a number of distinct irreducible representations at a given energy level. Such decomposition in terms of representations allows us to get further insights into the structure of the states and further information on the degeneracies. 

It is demonstrated that these representations share some similarities with those of $\mathfrak{su}(3)$ and that states of given energy can be decomposed into sum of irreducible representations which take the form of $\mathfrak{su}(3)$-like triangular multiplets.
In recent years, it was discovered how quadratic, cubic and more generally polynomial algebras appear in different contexts of mathematical physics \cite{pol05,Li2011,PDJRudolphYates2011,li14,yat18,alb18,vin20,cram20}. The class of polynomial algebras and related representations we have obtained in this paper are of interest for further applications from both mathematical and physical perspectives.

\subsection*{Acknowledgements}

\noindent This work was supported by the Future Fellowship FT180100099 and Discovery Project DP190101529 from the Australian Research Council.

\addcontentsline{toc}{chapter}{Bibliography}
\bibliographystyle{utphys}
\bibliography{Bibliography}

\end{document}